\documentclass[epj]{svjour}
\usepackage[utf8]{inputenc}
    \usepackage[T1]{fontenc}
\usepackage{latexsym}
\usepackage{graphics}
   \usepackage{amssymb}
   \usepackage{amsmath}
   \usepackage{amsbsy}
    \usepackage{bm}
\usepackage{graphicx}
\usepackage{color}
\begin{document}
\title{Dynamics of vortices and drift waves: a point vortex model}

\author{Xavier Leoncini\inst{1} \and Alberto Verga\inst{2}
}                     
\offprints{}          
\institute{Aix-Marseille Universit\'e, CPT-CNRS UMR 6207, Campus de Luminy, Case 907 - 13288 Marseille cedex 9, France \email{Xavier.Leoncini@cpt.univ-mrs.fr} \and Aix-Marseille Universit\'e, IM2NP-CNRS UMR 6242, 
   Campus de St J\'er\^ome, Case 142, 13397 Marseille, France \email{Alberto.Verga@univ-amu.fr}}
\date{Received: date / Revised version: date}
%
\abstract{
The complex interactions of localized vortices with waves is investigated using a model of point vortices in the presence of a transverse or longitudinal wave.  This simple model shows a rich dynamical behavior including oscillations of a dipole, splitting and merging of two like-circulation vortices, and chaos. The analytical and numerical results of this model have been found to predict under certain conditions, the behavior of more complex systems, such as the vortices of the Charney-Hasegawa-Mima equation, 
where the presence of waves strongly affects the evolution of large coherent structures.
} 
\maketitle
\section{Introduction}
\label{Sec:Intro}
Physical systems like large scale motion in oceans and atmospheres \cite{Zabusky82,Hasegawa79,zeitlin2007}, space and laboratory plasmas \cite{Hasegawa77,Diamond-2005fk}, or in statistical physics, the XY planar spin model \cite{KT73,leoncini98}, are all able to generate both vortices and waves, which interaction is one of the keys to their behavior. The phenomenology of these systems is very rich, ranging from turbulence to self-organization, formation of coherent structures \cite{McWilliams84,Provenzale99,Bracco2000,Reznik-2010fk} and anomalous transport \cite{Dupont98,Annibaldi02,Leoncini05}. Understanding the elementary mechanisms dominating the dynamics is crucial to model the statistical properties of these flows. One important fundamental process is the vortex interaction in presence of waves.
Besides complex dynamical patterns, particular behaviors have been identified, indeed, it has been observed that collisions of vortices may induce their merging, and on the other hand, the inverse process of splitting is also possible, large vortices may break under the action of waves. 
Both mechanisms are accompanied by the emission of waves, as for instance in the interaction of plasma vortices and drift waves \cite{Horton90,Fontan1995,Horton-1999fk}. 

The role played by waves in the interaction and evolution of vortices is not well understood. A comparison of the long-time evolution of Euler decaying two-dimensional turbulence \cite{Bracco2000} and the Charney-Hasegawa-Mima one (which supports wave modes) \cite{Matthaeus91}, shows that  the relaxation towards a kind of thermodynamic equilibrium state \cite{Onsager49} is much longer or unreachable in presence of a wave field \cite{Fontan1995,Agullo-2004rc,Spineanu-2005}. This illustrates that vortex interactions are largely influenced by the presence of waves.
 
A common feature of these vortex-wave systems is that they are 
 two-dimensional and may be described by a conservation equation which is a slight generalization of the vorticity Euler equation. In general the existence of vortex solutions is related to the so-called Poisson nonlinearity, which is the two-dimensional version of the convective nonlinearity in a normal fluid, or the electrostatic drift in magnetized plasmas. Besides, waves may be related to some source of divergence of the velocity field, as compressibility or as in the two-dimensional reduction of a stratified flow. The equation for the generalized vorticity $\Omega$, containing these basic mechanisms (a linear dispersion relation, and a Poisson nonlinearity) is given by
\begin{equation}
\frac{\partial\Omega}{\partial t}+ 
\left[\psi,\Omega\right]=0\:,
\label{vorti}
\end{equation}
where \([f,g]=f_xg_y-f_yg_x\) (\(f\) and \(g\) are arbitrary functions of \((x,y)\), and the subscripts denote derivatives) is the usual Poisson bracket, and $\psi$ is the stream function. The actual relation $\Omega = F(\psi)$ depends on the considered physical system, for the Euler equation it is simply given by $\Omega=-\nabla^2\psi$. This equation expresses the conservation of the generalized vorticity following the current lines. The Hasegawa-Mima equation for plasma drift waves \cite{Hasegawa77,Horton-1994}, also known in fluid mechanics as the Charney geostrophic equation for Rossby waves \cite{Charney-1948uq,Pedlosky_book,Lessieur_book_90}, reduces to (\ref{vorti}) with the generalized vorticity,
\begin{equation}
\Omega=-\nabla^2\psi+\psi/\rho_s^2-(v_d/\rho_s^2)\,x\:,
\label{mimeq}
\end{equation}
where $\psi$ is, in the plasma case, related to the electric potential(in suitable units), $\rho_s$ is the hybrid Larmor radius (Rossby length in the atmosphere) and $v_d$ is the drift wave velocity. The similarity of the equations governing Rossby waves in the atmosphere and drift waves in a plasma, may be traced back to the analogy between Coriolis and Lorentz forces, which determine the form of the dispersion relation, and the common convective nonlinearity, which reduces to a Poisson bracket in two dimensions. These two parameters, a length scale and a characteristic velocity, are related to important physical effects that modify the system's dynamics with respect to the one of the simpler Euler fluid. Indeed, the Charney-Hasegawa-Mima equation admits both waves and localized vortices solutions \cite{Horton-1999fk,Kim2002,Spineanu-2005zr,Reznik-1992ys,Reznik-2010fk}, allowing a rich nonlinear dynamics as shown in numerical simulations.  For instance, a dipole vortex has an oscillatory trajectory when its 
symmetry axis is inclined with respect to the wave direction of propagation \cite{Makino81}, or like-sign vortices (monopoles) may merge, under suitable conditions \cite{Horton90}.

 The drift velocity is related to the phase velocity of drift waves, the dispersion relation being 
\begin{equation}
\omega_{\bm k}=v_d k_y/(1+\rho_s^2 k^2)\,,
\label{dispersion}
\end{equation}
with $\omega$ the frequency and ${\bm k}$ the two-dimensional wavenumber. In the limit of vanishing drift velocity, localized vortices can become point vortices, 
\begin{equation}
\Omega\rightarrow \frac{\Gamma}{2\pi}\delta(\bm x)
\label{odelta}
\end{equation}
of circulation \(\Gamma\), with an interaction range of the order of the characteristic length \(\rho_s\).

The vortex-wave interaction is in general extremely complicated, involving the emission and scattering of waves by vortices and simultaneously the deformation of the vortex shape by the wave field \cite{Lansky-1997uq}. In this paper we investigate the interaction of localized vortices with waves, using the approximation that the wave is fixed, not affected by the vortices. We distinguish two situations according to the longitudinal or transverse character of the imposed wave. Transverse waves are naturally generated by the dynamics of Eq.~(\ref{vorti}), while longitudinal ones appear as a source term associated with the longitudinal component of the pressure driven velocity \cite{Lorenz-1960fk,Charney-1971vn}. We show that in both cases the interaction of  vortices (with same topological charge) with the wave can trigger their merging, or depending on the initial positions and wave strength, their separation. Moreover complex, chaotic or quasiperiodic, behavior can be found. We note as well that eventhoug in a different  context, 
similar features can be observed when ``perturbing''  the integrable motion of two equal-sign vortices with noise \cite{Agullo97,sire2011}.

In the following section, Sec.\ \ref{Sec:Basic}, we state the basic equations of the point vortex model for the transverse and longitudinal external wave, and discuss its relation with Charney-Hasegawa-Mima equation. In Secs.~\ref{Sec:PVt} and \ref{Sec:PVl} we investigate the different dynamical regimes that the presence of the wave, transverse and longitudinal respectively, can induce in the motion of the point vortices. We describe the dipole (opposite circulation vortices) and the monopole (like-circulation vortices) trajectories. We find the range of parameters where the strong interaction of the monopole and the wave leads to their merging or splitting. In Sec.~\ref{Sec:HM} we compare the results of Secs.~\ref{Sec:PVt} and \ref{Sec:PVl}, with numerical simulations of the Charney-Hasegawa-Mima equation, and finally (Sec.~\ref{Sec:Con}) we conclude with a summary and brief discussion of the results.

\section{Transverse and longitudinal waves: point vortex model}
\label{Sec:Basic}
In principle, the wave term in the generalized vorticity (\ref{mimeq}), being proportional to the $x$-coordinate and obviously non localized, forbids a solution  in the form of an assembly of point vortices \cite{Kono-1991fk}. However, as a first approximation, we may consider a system where the vorticity is highly concentrated but subject to the action of a wave field in such a way that the effect of the vortices on the wave may be neglected. If one neglects the internal structure of the localized vortices and concomitantly, the feedback effects on the wave, the velocity field can be considered as resulting from the superposition of two terms, one related to the point vortices and one related to the wave. Therefore, our basic model will be two point vortices driven by the action of an external wave. It mimics the phenomenology of the Charney-Hasegawa-Mima system, dominated by the dynamics of coherent structures evolving in the field of transverse and driven longitudinal waves. We 
show that, in spite of its extreme simplicity, this system contains a rich dynamics that can model processes such as splitting and fusion, and deterministic or chaotic propagation of vortices. 

Point vortices are defined by a potential vorticity distribution given by a superposition of Dirac functions,
\begin{equation}
\Omega_l({\bm x},t)=
   \frac{1}{2\pi}\sum_{\alpha=1}^N\Gamma_{\alpha}
   \delta\left({\bm x}- {\bm x}_{\alpha}(t)\right)\,,
\label{omegal}
\end{equation}
where ${\bm x}$ is a vector in the plane of the flow, $\Gamma_{\alpha}$ is the circulation of vortex $\alpha$, $N$ is the total number of vortices, and ${\bm x}_{\alpha}(t)$ is the vortex position at time $t$. Point vortices are known to be exact solutions of Euler equation \cite{Machioro94}, that is when the vorticity is given by 
\begin{equation}
\Omega=-\nabla^2\psi_l\:,
\label{omega_euler}
\end{equation}
 in terms of the local stream function \(\psi_l\), but they are also exact solutions of the more general equation (\ref{vorti}), for 
\begin{equation}
 \Omega=\Omega_l=-\nabla^2\psi_l+\psi_l/\rho_s^2\:.
\label{omega_Lamb}
\end{equation}
 The usual \(\psi_l\) (solution of the Poisson equation), which gives a logarithmic interaction, is replaced in the second case (\ref{omega_Lamb}), by a modified Bessel function ${\rm K}_0$ interaction (solution of the Helmholtz equation). 

The presence of the wave should add to (\ref{omegal}) a regular source of potential vorticity. Considering that the corresponding stream function is the superposition of a localized (vortex) and a wave term, \(\psi=\psi_v+\psi_w\), the motion of vortices will be governed by,
\begin{equation}
\dot{\bm x}_\alpha = \bm v(\bm x,t)|_{\bm x = \bm x_\alpha},\quad
	\bm v=\hat{\bm z}\times \nabla(\psi_v+\psi_w)+\nabla \phi
\label{vadvection}
\end{equation}
where the velocity field is calculated at the position of the vortex \(\alpha\), excluding the singular contribution of the \(\alpha\) vortex and $\hat{\bm z}$ denotes the unit vector parallel to the $z-$axis. In addition to the transverse wave, the last term \(\phi\), takes into account the possibility of a longitudinal wave driven by an external source. 

In order to explicitly display the approximation underlying the point vortex model of Eq.~(\ref{vadvection}), we return to the Charney-Hasegawa-Mima equation (\ref{vorti}), that can be written as
\begin{equation}
\frac{\partial}{\partial t}\Omega_l
   + \frac{v_d}{\rho_s^2}\frac{\partial}{\partial y}\psi + 
   [\psi,\Omega_l]=0\,,
\label{CHM}
\end{equation}
where $\Omega_l$ given by Eq.~(\ref{omega_Lamb}),
%
%
is the localized potential vorticity. Introducing the decomposition \(\psi=\psi_v+\psi_w\), one obtains,
\begin{align}
\frac{\partial}{\partial t}(-\nabla^2\psi_v + \psi_v/\rho_s^2) +
	[\psi_v+\psi_w,-\nabla^2\psi_v + \psi_v/\rho_s^2]&	
\nonumber\\
	+ \frac{\partial}{\partial t}(-\nabla^2\psi_w + \psi_w/\rho_s^2)
    + \frac{v_d}{\rho_s^2}\frac{\partial}{\partial y}\psi_w + 
    [\psi_w,-\nabla^2\psi_w]&	
\nonumber\\
    + \frac{v_d}{\rho_s^2}\frac{\partial}{\partial y}\psi_v + 
    [\psi_v,-\nabla^2\psi_w]=0 &\,.
\label{CHM-vw}
\end{align}
We consider now that the localized component of the potential vorticity, $\Omega_l=\Omega_l(\psi_v)$, is given by the superposition of point vortices  (\ref{omegal}). Then, the first line of (\ref{CHM-vw}) only contains $\delta$-localized terms, and vanishes if the point vortices trajectories are precisely given by (\ref{vadvection}) (without the source term, in \(\nabla\phi\)). Therefore, the point vortex-wave model  requires the fulfilling of the following two conditions. First, it assumes that the terms on the last line of (\ref{CHM-vw}), that is the dispersion effect on the vortices and the influence of the vortex on the wave, are negligible. Second, we remark that for a single drift wave (a wave satisfying the linear dispersion relation) the second line vanishes identically.


In general (\ref{CHM}) is complemented with a source and dissipation terms, that we write as,
\begin{equation}
\frac{\partial}{\partial t}\Omega_l
   + \frac{v_d}{\rho_s^2}\frac{\partial}{\partial y}\psi + 
   [\psi,\Omega_l]=\nu \nabla^4\psi + S(\bm x,t)\,,
\label{CHMs}
\end{equation}
where \(\nu\) is the kinematic viscosity, and \(S(\bm x,t)\) an external vorticity source. In the Appendix~\ref{Sec:app}, we show that the source term can be naturally associated with longitudinal wave perturbations.

In summary, we apply the velocity field derived from the stream function whose dynamics is governed by (\ref{CHM}) or (\ref{CHMs}), to the point vortex model (\ref{vadvection}). In the source free case (\ref{CHM}), the velocity field is purely transverse, and we may write it as,
\begin{equation}
\bm v(\bm x, t)=\hat{\bm z}\times\nabla\psi(\bm x,t)\,,\quad
\psi=\psi_l+\psi_w
\label{vt}
\end{equation}
where the stream function is assumed to result from the superposition of the point vortex term \(\psi_l\) and the nonlocal (wave) term \(\psi_w\). In contrast, in the forced case the velocity field may not be transverse, and contain a longitudinal (potential) contribution. For the driven case (\ref{CHMs}), the velocity field can then be chosen as,
\begin{equation}
\bm v(\bm x, t)=\hat{\bm z}\times\nabla\psi_l(\bm x,t) + \nabla \phi(\bm x, t)\,,
\label{vl}
\end{equation}
in accordance with (\ref{vtotal}) but with \(z\)-independent quantities, where, in addition to the rotational component derived from the stream function \(\psi_l\), there is a gradient component, whose physical origin is related to the non-stationary pressure perturbations in the vertical direction \cite{Pedlosky_book,Lorenz-1960fk,Charney-1971vn}. The first case (\ref{vt}) corresponds to a system of point vortices in the field of a transverse wave, and the second case (\ref{vl}), to the presence of a longitudinal wave.

\begin{figure} 
\centering
\includegraphics[width=0.45\textwidth]{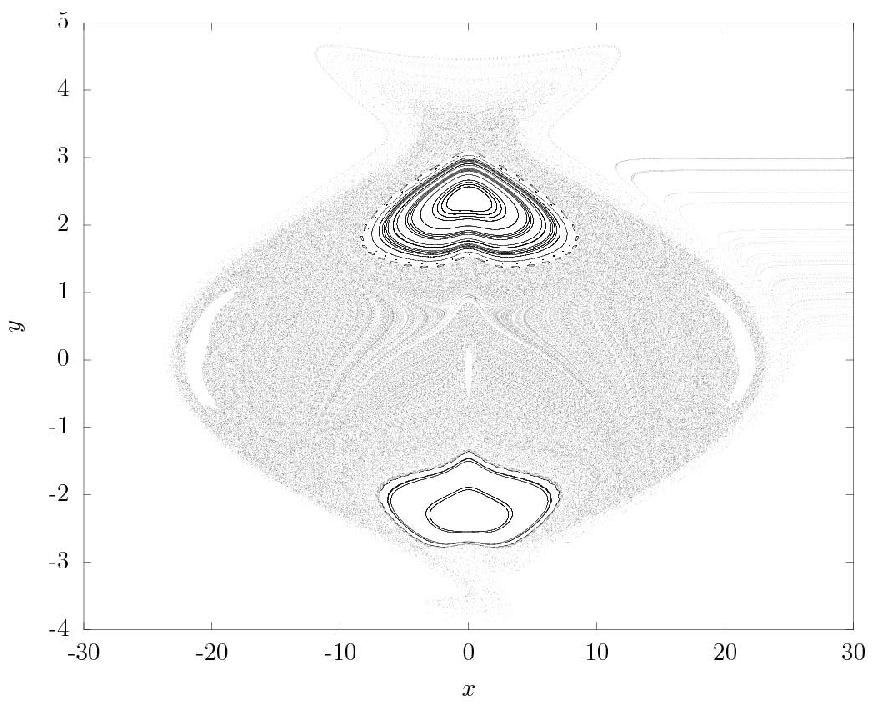}(a)
\includegraphics[width=0.45\textwidth]{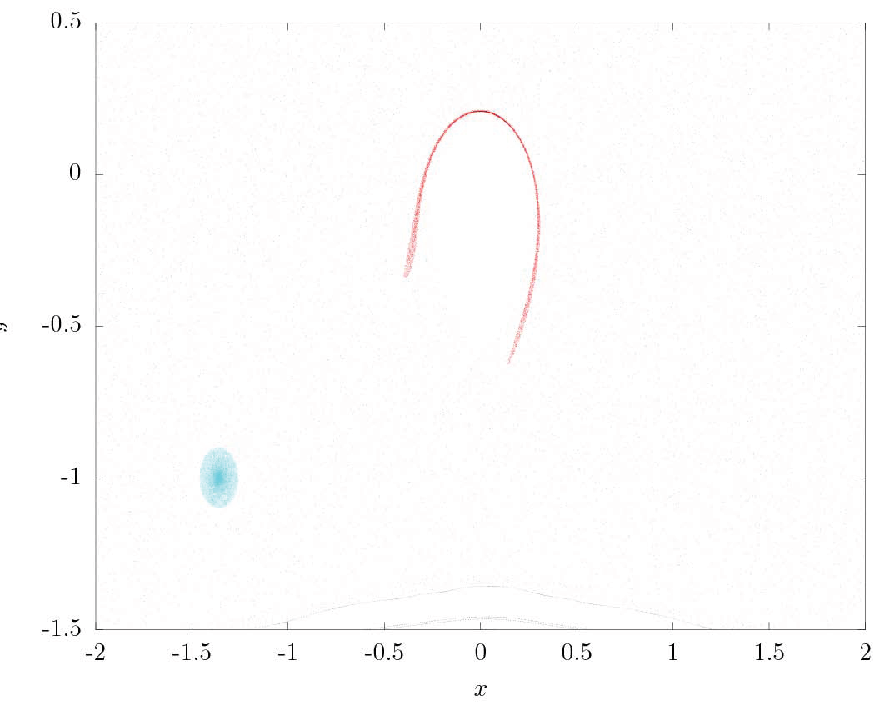}(b)
\caption{(Color online) Poincaré map of the two vortices-wave system (\ref{tnum}), in $(x=x_1-x_2,y=y_1-y_2)$ phase plane (a); runaway trajectories (vortex splitting) are present together with others approaching the origin (vortex merging), and large chaos regions. (b) Zoom of the central region that show ``merging'' trajectories, an ensemble of initial points (blue circle) approaches the center (red arc) in one period.}
\label{f-poincare1}
\end{figure}

The motion of the vortices, as a consequence of the Helmholtz theorem \cite{Saffman95}, is determined by the value of the velocity \(\bm v=\hat{\bm z}\times\nabla\psi+\nabla \phi\) [which contains both cases (\ref{vt}) and (\ref{vl})], at the position of the vortices, as already shown in (\ref{vadvection}). If the driven potential \(\phi\) vanishes, this motion is Hamiltonian. In the general case the point vortex equations can be written as,
\begin{equation}
\Gamma_{\alpha} \dot{\bm x}_{\alpha}=
   \hat{\bm z}\times\frac{\partial H}{\partial {\bm x}_{\alpha}}+
   \Gamma_{\alpha}\nabla\phi(\bm x_\alpha,t)\,.
\label{pvor+}
\end{equation}
where the Hamiltonian $H$ is given by
\begin{equation}
H=\frac{1}{2\pi}
  \sum_{\alpha>\beta}\Gamma_{\alpha}\Gamma_{\beta}
  U(|{\bm x}_{\alpha}-{\bm x}_{\beta}|)+\sum_{\alpha}\Gamma_{\alpha}\psi_w(\bm x_\alpha,t)\,,
\label{ham}
\end{equation}
with $U(x)=-\log(x)$ in the Euler case, and $U(x)={\rm K}_0(x/\rho_s)$, the modified Bessel function of zero order, in the more general case (when $\rho_s\rightarrow\infty$ it tends to the logarithm). When the distance between the vortices is smaller than the typical interaction length $\rho_s$ the behavior of the two systems is similar, in the opposite case, the ${\rm K}_0$ interaction decreases exponentially and the vortices become ``almost free''.

The Hamiltonian (\ref{ham}) explicitly depends on time, and in the case of two vortices, reduces to the class of one and a half degrees of freedom, which generically display Hamiltonian chaos \cite{ZaslavBook98}. We can therefore expect that  two initially well separated trajectories may approach each other closely or, at variance, two initially close vortices may follow diverging trajectories. Translating these behaviors to the extended system described by the Charney-Hasegawa-Mima equation, we anticipate  merging and splitting of localized structures.

\section{Dynamics of two point vortices in the field of a transverse wave}
\label{Sec:PVt}
We assume that a particular solution of (\ref{CHM}), as discussed in the previous section, can be approximated by \(\psi=\psi_l+\psi_w\), where the first term represents two point vortices of circulations \(\Gamma_1\) and \(\Gamma_2\), and the second term, a wave \(\psi_w=\Gamma_0 \cos(ky-\omega t)\) of amplitude \(\Gamma_0\), wavenumber \(\bm k=(0,k)\), and frequency \(\omega\). The motion equation (\ref{pvor+}) of the vortex pair writes,
\begin{subequations}
\label{tnum}
\begin{align}
\dot{\bm{x}}_1&=\frac{\Gamma_2}{2\pi\rho_s}\left(
	\begin{array}{c} y_1-y_2\\ 
		x_2-x_1\end{array}\right)
	\frac{\mathrm{K}_1(r/\rho_s)}{r}-\hat{\bm{x}}\Gamma_0k\sin(ky_1-\omega t)\,,\\
\dot{\bm{x}}_2&=\frac{\Gamma_1}{2\pi\rho_s}\left(
	\begin{array}{c} y_2-y_1\\ 
		x_1-x_2\end{array}\right)
	\frac{\mathrm{K}_1(r/\rho_s)}{r}-\hat{\bm{x}}\Gamma_0k\sin(ky_2-\omega t)\,,
\end{align}
\end{subequations}
where \(\bm x_i=(x_i,y_i)\), \(i=1,\,2\), are the vortex positions, \(r\) their distance of separation, \(\mathrm{K}_1\) the modified Bessel function of first order, and $\hat{\bm x}$ the unit vector parallel to the $x-$axis. In the absence of wave, the system behaves as a nonlinear oscillator whose frequency is \((\Gamma_1 + \Gamma_2)\,\mathrm{K}_1(r_0/\rho_s)/4\pi\rho_s r_0\) (where \(r_0\) is the initial distance between the two vortices). However, driven by the wave, the dynamics of this oscillator will generically exhibit chaotic trajectories.

In the case of two identical point vortices of circulation \(\Gamma=\Gamma_1=\Gamma_2\), the center of mass motion separates from the relative motion \cite{remark}. Using the coordinates \((x,y)=(x_1-x_2,y_1-y_2)/2\) and \((X,Y)=(x_1+x_2,y_1+y_2)/2\), the relative motion is described by the set,
\begin{subequations}
\label{relative}
\begin{align}
\dot{x}&=\frac{\Gamma y}{2\pi\rho_s r}
	\mathrm{K}_1\left(\frac{r}{\rho_s}\right)-
	\Gamma_0k\cos(kY-\omega t)\sin ky\,,\\
\dot{y}&=-\frac{\Gamma x}{2\pi\rho_s r}
	\mathrm{K}_1\left(\frac{r}{\rho_s}\right)\,,
\end{align}
\end{subequations}
where \(r=2(x^2+y^2)^{1/2}\), and \(Y=\mathrm{const}\). The important point about these equations is that they can be derived from an effective Hamiltonian:
\begin{equation}
H_T=\frac{\Gamma}{2\pi}
	\mathrm{K}_0\left(\frac{r}{\rho_s}\right)+
	\Gamma_0\cos(kY-\omega t)\cos ky
\label{heff}
\end{equation}
which describes the one and a half degrees of freedom dynamics in the \((x,y)\) phase space, and constitute then a convenient representation to introduce a Poincaré map, using the natural period \(2\pi/\omega\) of the driven wave.

To identify the diverse dynamical patterns of (\ref{tnum}), corresponding to a set of initial conditions, we represent in Fig.~\ref{f-poincare1} the Poincaré map of the phase space \(\bm r=(x,y)\) using the parameters \(\Gamma = 8\pi\), \(\Gamma_0=5.6\), \(k=3\pi/10\) and \(\omega=k/(1+k^2)\)  (in units of \(\rho_s\) and \(v_d\)). We may distinguish periodic and quasiperiodic, chaotic and runaway trajectories, depending on the initial positions of the vortices. Chaotic trajectories may include a set of initial conditions that rapidly (in a few periods) approaches the origin, as shown in Fig.~\ref{f-poincare1}(b). This particular kind of trajectories, in an extended dissipative system, should correspond to merging of vortices. Moreover, runaway trajectories should correspond to splitting of initially bounded monopoles. We show in Sec.~\ref{Sec:HM}, that these processes, including complex monopole interactions, are effectively observed in the dynamics of the Charney-Hasegawa-Mima equation.

\begin{figure}
\includegraphics[width=0.45\textwidth]{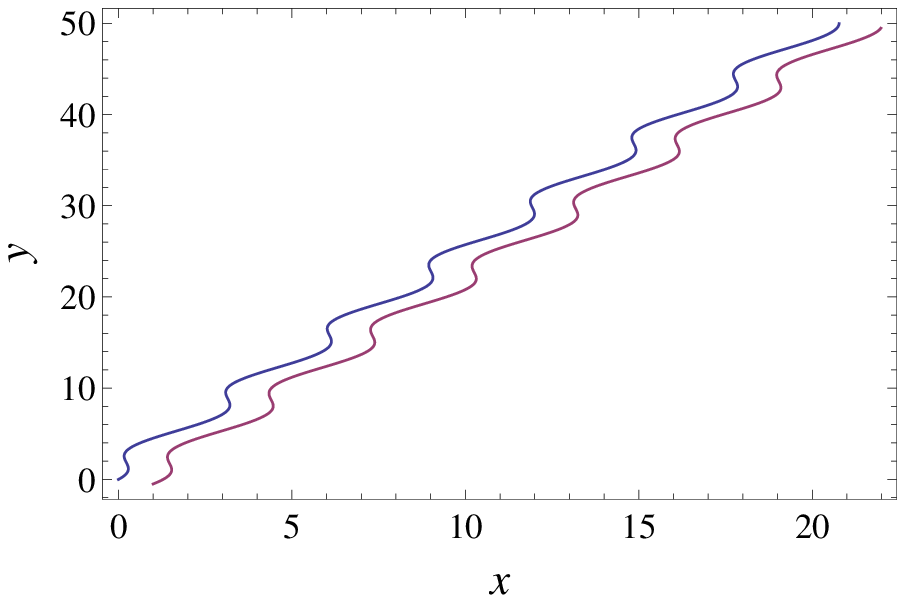}(a)\hfill%
\includegraphics[width=0.45\textwidth]{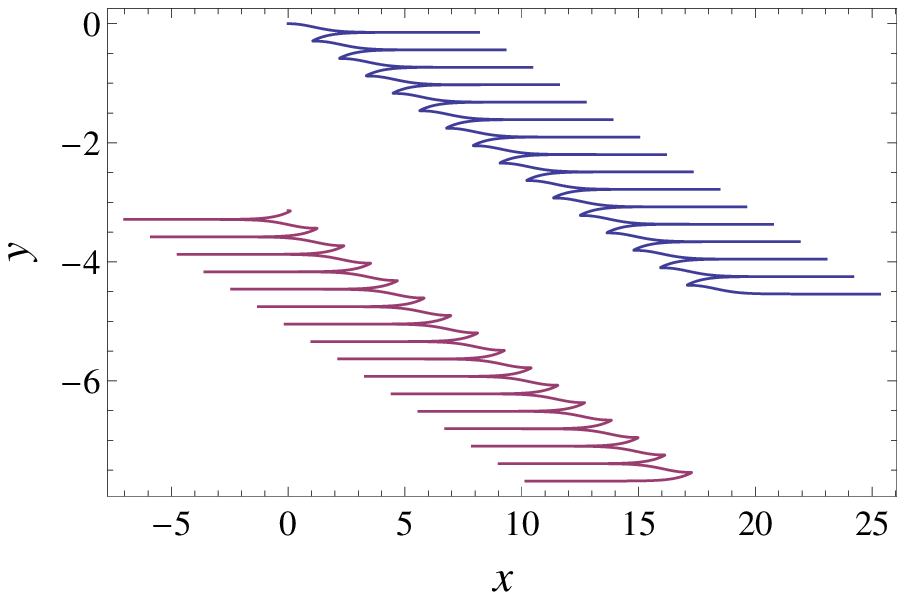}(b)
\caption{(Color online) Trajectory of the dipole vortex in a transverse wave. Parameters:  $g=8\pi$,
$a=4$, $\bm{x}_1=(0,0)$ (a) A representative case of the oscillatory dipole motion $\bm{x}_2=(1,-0.5)$; (b) particular solution showing horizontal excursions $\bm{x}_2=(0,-\pi)$.}
\label{f-dt}
\end{figure}

If the two vortices have opposite circulations \(\Gamma_1=-\Gamma_2\), the system (\ref{tnum}) becomes integrable, 
\begin{subequations}
\label{tdip} 
\begin{eqnarray}
\label{tdip-x}
	\dot x&=&-\Gamma_0 k\sin ky\cos(kY-\omega t)\,,\\
\label{tdip-y}
	\dot y&=&0 \,,\\
\label{tdip-X}
	\dot X&=&\frac{\Gamma y}{2\pi\rho_s}\frac{\mathrm{K}_1(r/\rho_s)}{r} -
	\Gamma_0 k \cos ky\sin(kY-\omega t)\,,\\
\label{tdip-Y}
	\dot Y&=&-\frac{\Gamma x}{2\pi\rho_s}\frac{\mathrm{K}_1(r/\rho_s)}{r} \,,
\end{eqnarray}
\end{subequations}
from which one finds that \(y=\mathrm{const.}\), and \(r=r(t)\) separates from the center of mass equations. Hence, the solutions of these equations are typically dipole trajectories that follow a straight line to which is superposed an oscillation (Fig.~\ref{f-dt}a). One particular solution is obtained in the case \(y=n\pi/k\) with \(n\) integer, the vortex separation in the \(y\) direction is a multiple of the wavelength. In this case the vortices drift with a constant velocity in the \(y\) direction and execute horizontal excursions in the form,
\begin{equation}
X(t)=V_0t-\Gamma_0 k \cos(t/t_0)\,,
\end{equation}
where the constants \(V_0\) and \(t_0\) depend on the initial conditions. These horizontal excursions are present in the trajectory shown in Fig.~\ref{f-dt}b.

\section{Dynamics of two point vortices in the field of a longitudinal wave}
\label{Sec:PVl}
We consider now the motion of two vortices in the presence of a longitudinal wave. The basic equations of the model are directly derived from
(\ref{pvor+}), where we put \(\psi_w=0\) and choose a wave potential simply defined by
$\phi=\Gamma_0\cos({\bm k}\cdot{\bm x}-\omega t)$, with $\Gamma_0$
the wave amplitude, ${\bm k}$ the wave vector, and $\omega$ the
wave frequency,
\begin{equation}
\dot{\bm r}=
  \Gamma_{\!+}\hat{\bm z}\times\frac{{\bm r}}{r^2}-
  2\Gamma_0 {\bm k}
  \cos\left({\bm k}\cdot{\bm R}-\omega t\right)
  \sin\left(\frac{{\bm k}\cdot{\bm r}}{2}\right)\; ,
\label{dyn1}
\end {equation}
\begin{equation}
\dot{\bm R}=
  \Gamma_{\!-} \hat{\bm z}\times \frac {{\bm r}}{2r^2}-
  \Gamma_0 {\bm k}
  \sin\left({\bm k}\cdot{\bm R} -\omega t\right) 
  \cos\left(\frac{{\bm k}\cdot{\bm r}}{2}\right)\; ,
\label{dyn2}
\end {equation}
(in the \(\rho_s\rightarrow\infty\) limit), where ${\bm r}={\bm r}_1-{\bm r}_2$ is the relative position of the
vortices, ${\bm R} =({\bm r}_1+{\bm r}_2)/2$ is their ``center of
mass'', and $\Gamma_{\!+}=(\Gamma_1+\Gamma_2)/2\pi$,
$\Gamma_{\!-}=(\Gamma_1-\Gamma_2)/2\pi$, are their reduced
circulations. For the
analytical computations we take the simpler
logarithmic interaction, although generalization to the modified Bessel interaction, should be straightforward.

\begin{figure}
\includegraphics[width=0.45\textwidth]{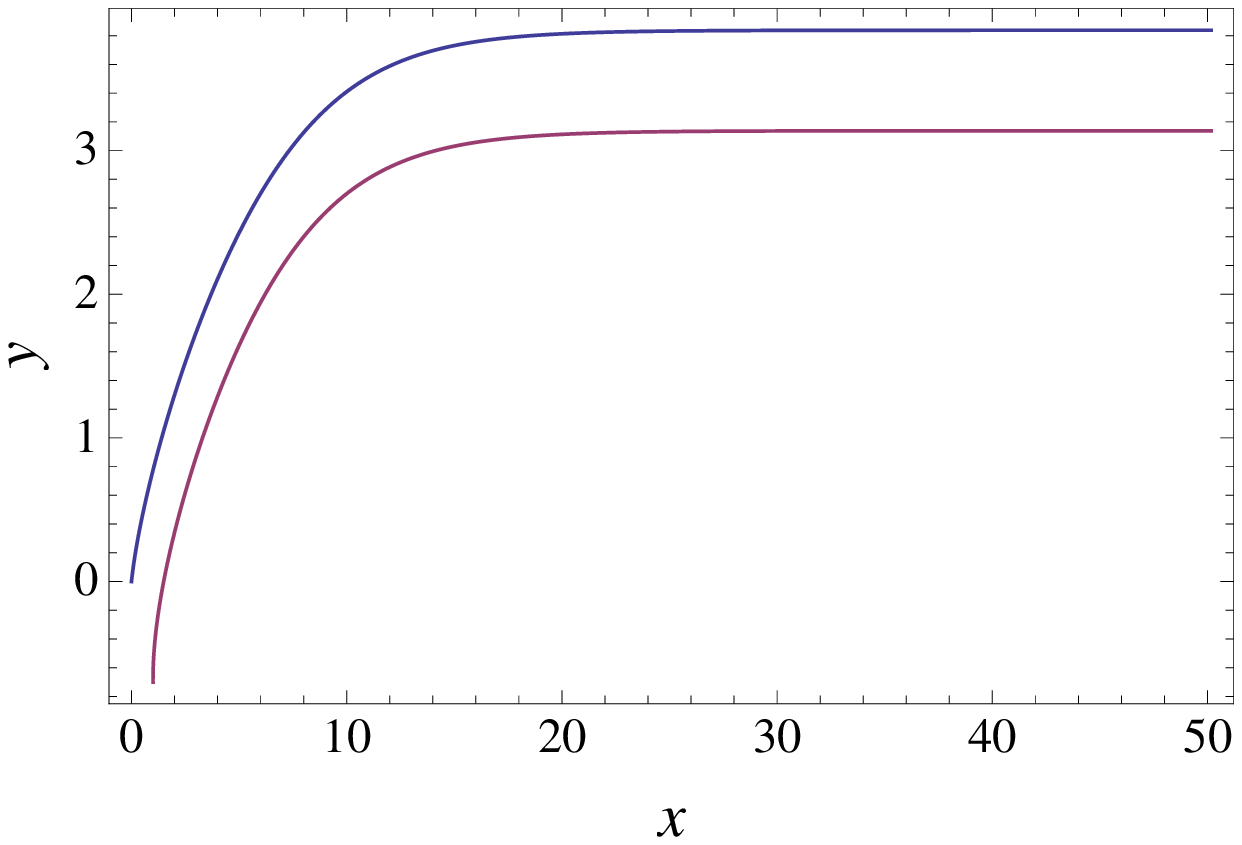}(a)\hfill%
\includegraphics[width=0.45\textwidth]{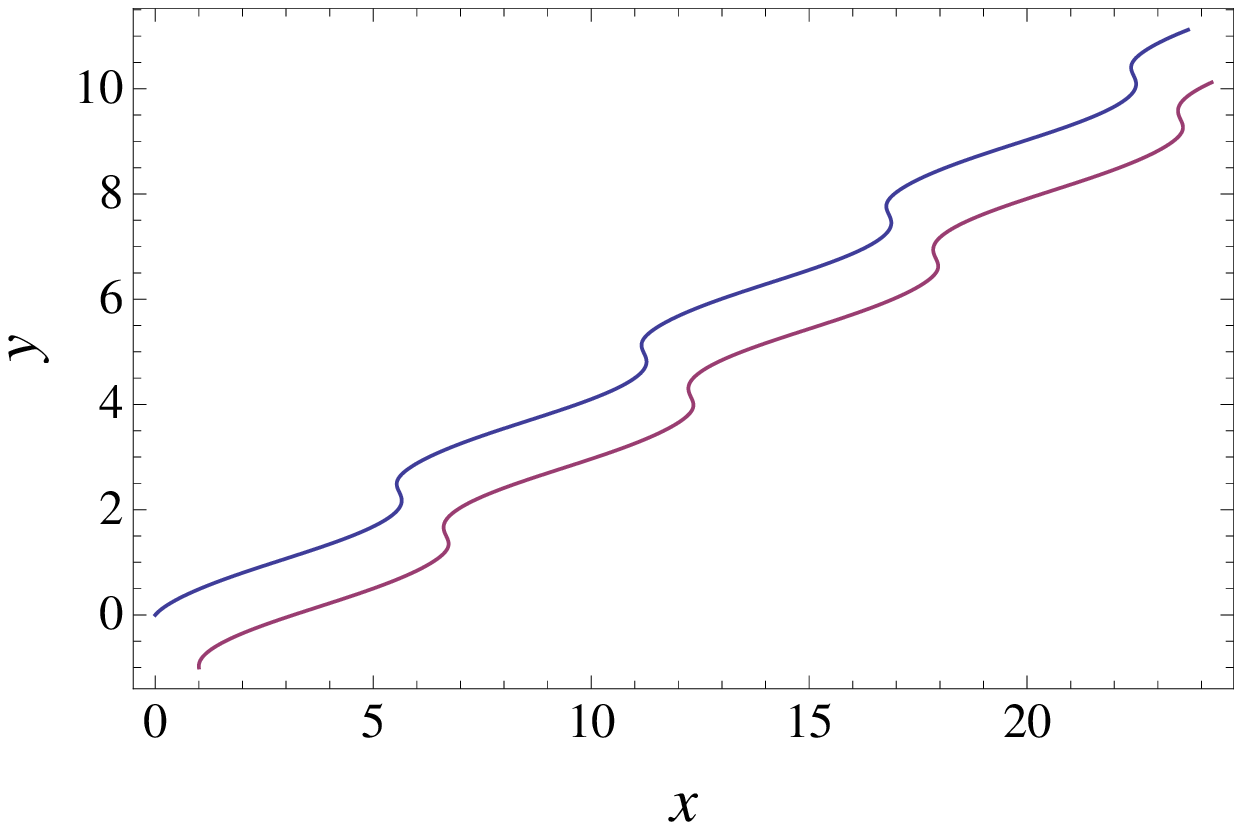}(b)
\caption{(Color online) Trajectory of the dipole vortex in a longitudinal wave. (a) A representative case of the dipole motion locking when \(|V|<\alpha_0\); (b) the weak wave case (\(|V|>\alpha_0\)), the dipole trajectory is modulated by the wave. Parameters:  $g=1$,
$a=0.3$, $\bm{x}_1=(0,0)$, $\bm{x}_2=(1,-0.7)$ (a) and $\bm{x}_2=(1,-1)$ (b), $t_f=50$.}
\label{f-dip}
\end{figure}

For further calculations, it is useful to measure lengths in $1/k$ 
and time in $1/\omega$ units, and to perform a Galilean transformation 
to the frame moving at the wave phase velocity, leading to the 
following non-dimensional variables and parameters:
 $x' = {\bm k}\cdot{\bm r}/2$, 
 $y' = {\bm k}\cdot(\hat{\bm z}\times{\bm r})/2$, 
 $t'=\omega t$, 
 $X' = {\bm k}\cdot{\bm R}-t$, 
 $Y' = {\bm k}\cdot(\hat{\bm z}\times{\bm R})$, the distance 
 ${\bm r'}=(x',y')$, and center of mass
 ${\bm R'}=(X',Y')$ vectors,
 and parameters
 $\alpha_0=\Gamma_0 k^2/\omega$, for the wave amplitude
 $\alpha_{\!+}=\Gamma_{\!+} k^2/4\omega$,
 $\alpha_{\!-}=\Gamma_{\!-} k^2/4\omega$, for the vortices circulation (in the following we drop the primes). 
The coordinate system is chosen, without loss of generality, so that
the $x$-axis is parallel to the vector ${\bm k}$.

In addition to the analytical calculations, we compute the point vortices trajectories in the presence of a longitudinal wave, by direct integration of the motion equations in the form,
\begin{subequations}
\label{num}
\begin{eqnarray}
\dot{\bm{x}}_1&=&g\left(
	\begin{array}{c} y_1-y_2\\ 
		x_2-x_1\end{array}\right)
	\frac{\mathrm{K}_1(\nu r)}{r}-\hat{\bm{x}}a\sin(x_1-t)\,,\\
\dot{\bm{x}}_2&=&\mp g\left(
	\begin{array}{c} y_2-y_1\\ 
		x_1-x_2\end{array}\right)
	\frac{\mathrm{K}_1(\nu r)}{r}-\hat{\bm{x}}a\sin(x_2-t)\,,
\end{eqnarray}
\end{subequations}
for arbitrary $\nu=1/k\rho_s$. We used $g=k\Gamma/\omega \rho_s$, where $\Gamma=\Gamma_1=-\Gamma_2$ for the dipole (minus sign), and $\Gamma=\Gamma_1=\Gamma_2$ for the monopole (plus sign), and $a=k\Gamma_0/\omega\rho_s$ (in units of $k=1$ and $\omega=1$). The vortex motions are qualitatively similar for finite range and logarithmic interactions; we show in the following, results for \(\rho_s=1\), and different choices of the initial positions and strengths \(g\) and \(a\).

Two cases are of special interest, the so called ``dipole'', when the vortices have opposite circulations 
$\alpha_{\!+}=0$, and the ``monopole'', when the vortices have the same circulation $\alpha_{\!-}=0$. Without wave the dipole moves in a straight line, and the monopole follows a circular trajectory. 

In the dipole case, the equations of motion reduce to the completely integrable system
\begin{subequations}
\label{dip} 
\begin{eqnarray}
\label{dip-x}
	\dot x&=&-\alpha_0\cos X\sin x\,,\\
\label{dip-y}
	\dot y&=&0 \,,\\
\label{dip-X}
	\dot X&=&-1-\alpha_0\,\sin X \cos x+\alpha_{\!-}\frac{y}{r^2}\,,\\
\label{dip-Y}
	\dot Y&=&-\alpha_{\!-}\frac{x}{r^2}\,.
\end{eqnarray}
\end{subequations}
where $r^2=x^2+y^2$. Examples of dipole trajectories are presented in Fig.~\ref{f-dip}.

\begin{figure}  
\centering
\includegraphics[width=0.45\textwidth]{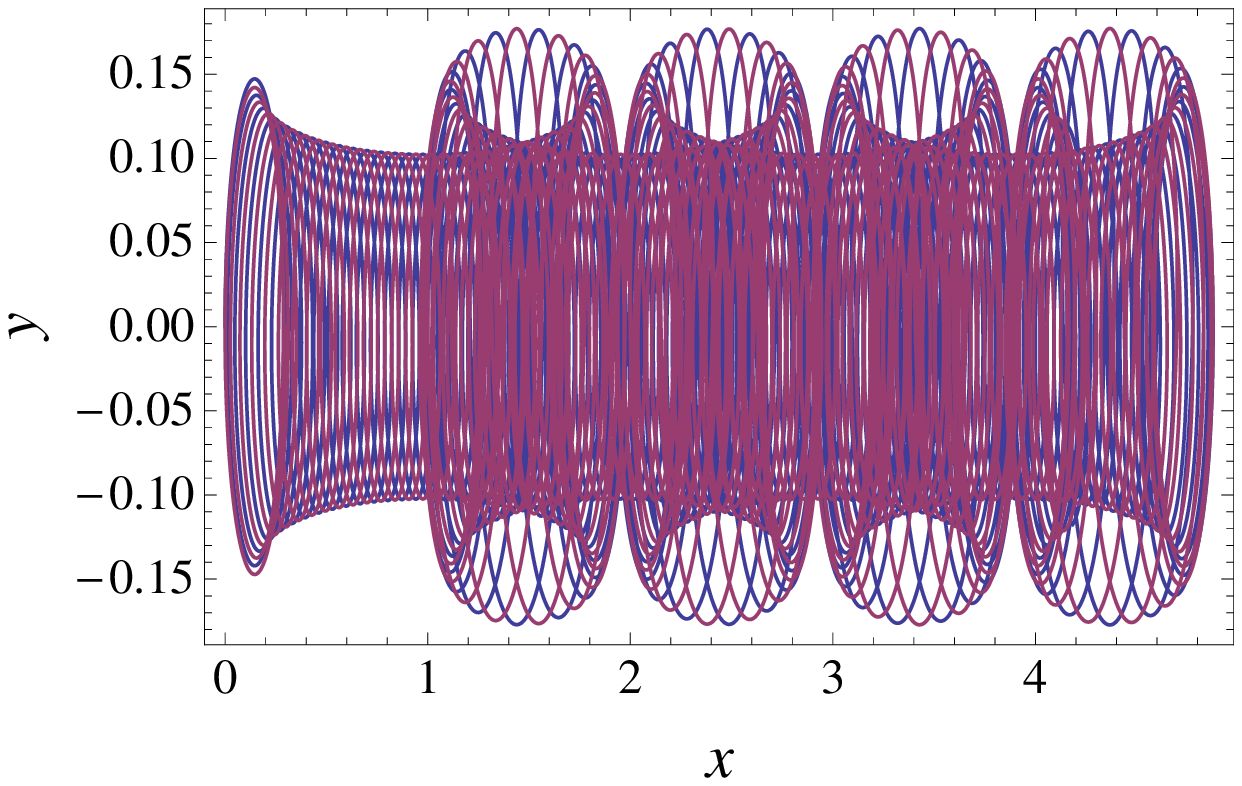}(a)\hfill%
\includegraphics[width=0.45\textwidth]{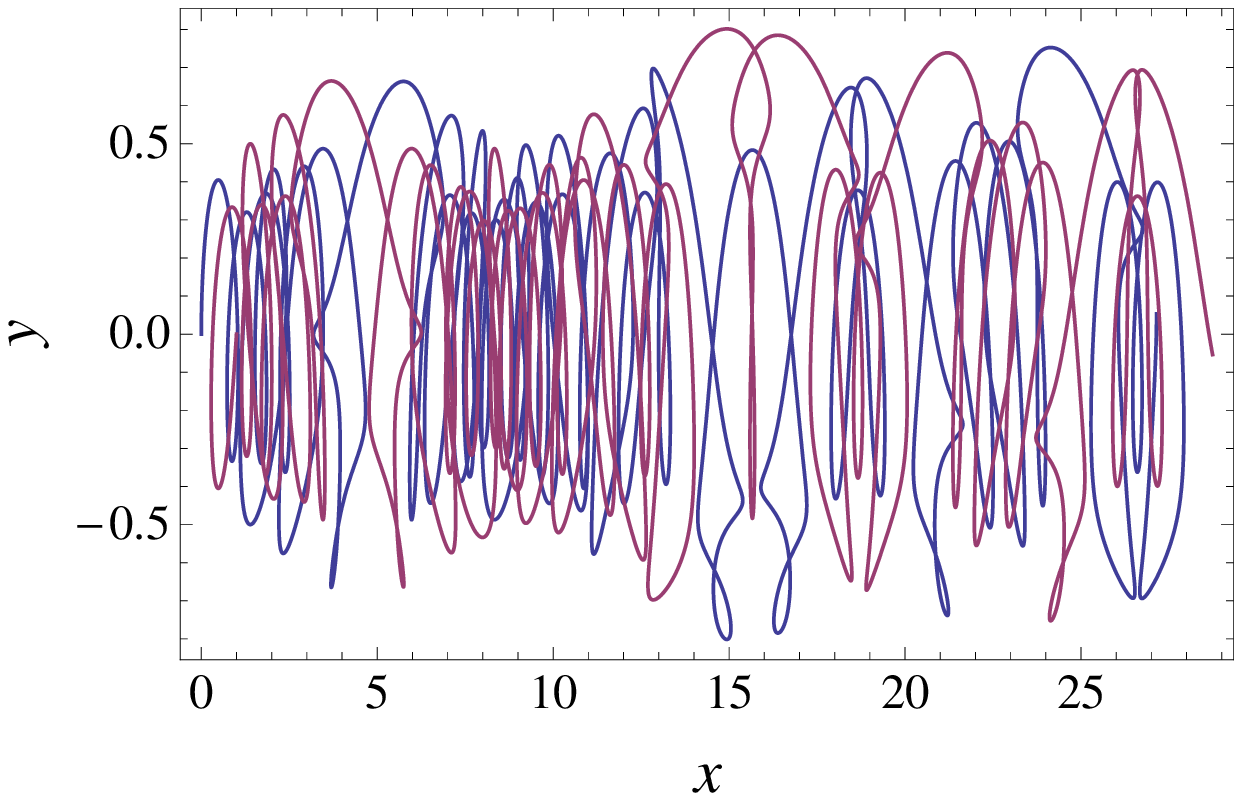}(b)
\caption{(Color online) Quasi-periodic (a) and chaotic (b) motion of the monopole, in the long-wave case. Parameters
 ${\bm x}_1=(0,0)$, ${\bm x}_2=(0.3,0)$ (a), and ${\bm x}_2=(1,0)$ (b), $t_f=200$, $\Gamma_1=1$, $\Gamma_2=1$, $\Gamma_0=0.5$,
$\omega=1$.
}
\label{f-qper}
\end{figure}

Although the motion of a dipole interacting with a wave is
integrable, to find a quadrature is difficult, so some insight can be
gained by studying particular exact solutions of (\ref{dip}). In analogy with the transverse wave case, solutions with ${\bm{r}}={\bm{r}}_0$ constant exist if $x=n\pi$ (the vortex distance in the \(x\) direction is a multiple of the wavelength). In such a case, we can define $\omega_0=\alpha_{\!-}/({r_0}^2)$, the characteristic
frequency of the dipole, and $V=-1+\omega_0 y_0$, the $x-$component
of the speed of the unperturbed dipole in the frame of the wave. We
obtain for ${\bm R}$
\begin{equation}
\dot X=V+(-1)^{n+1}\alpha_0\sin X
\hspace{10mm}
Y=-\omega_0 x_0 t\,.
\label{exotic}
\end{equation}
This equation naturally leads to two different cases, depending on
the value of $|V/\alpha_0|$, which determines the existence of a fixed
point for $X$. In one case ($|V|>\alpha_0|$) the $x$-component of the
dipole speed $\dot X$ has a constant sign, otherwise it changes sign  ($|V|<\alpha_0$) and thus can eventually approach
a fixed point.

%
We first notice, that in the very simple case where $V=0$ (the
$x$-component of speed of the unperturbed dipole is equal to the
phase velocity), which we call the resonant case, the equation
(\ref{exotic}) becomes $\dot X=(-1)^{n+1}\alpha_0 \sin X$. This
equation predicts the existence of selected directions $X=n\pi$ for
the dipole propagation, a feature reminiscent to the mode locking phenomenon as for instance the one observed by in \cite{paoletti2005experimental,paoletti2005front}. These directions correspond to the case where
the dipole is located at an extremum of the wave, and therefore where
the profile of the wave is flat. Therefore, if the condition $V=0$ is
satisfied the dipole will only see this flat profile and the influence
of the wave on the dipole motion vanishes. Indeed, the general
solution of (\ref{exotic}) for $V=0$ is
\begin{equation}
X=2\arctan\left(
\mathrm{e}^{\left(-1\right)^{n+1}\alpha_0 t}
\tan\frac{X_0}{2}\right)\,,
\end{equation}
and the dipole sets itself in one of the selected directions where it
does not see the wave.

\begin{figure} 
\centering
\includegraphics[width=0.45\textwidth]{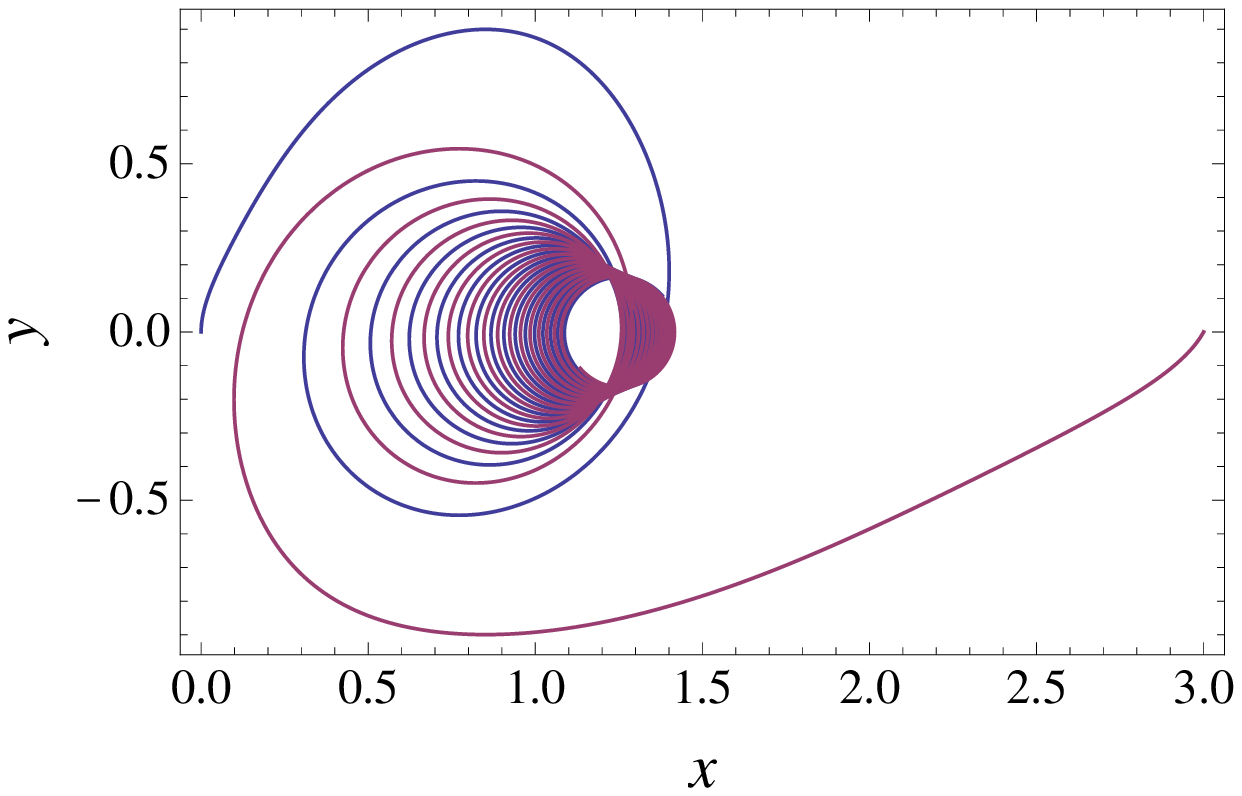}(a)\hfill%
\includegraphics[width=0.45\textwidth]{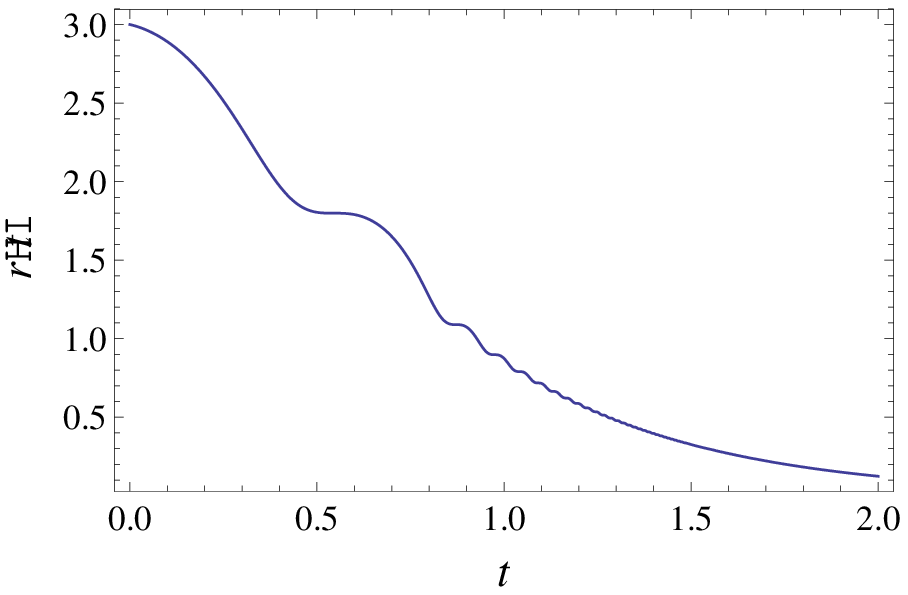}(a)
\caption{(Color online) Vortex trajectory leading to fusion. Typical motion in the $\alpha_0>1$ case; (a) trajectory $t\in(0,1.5)$, (b) vortex separation as a function of time. Parameters:
$\bm{x}_1=(0,0)$, $\bm{x}_2=(3,0)$, $\Gamma=8\pi$,
$\Gamma_0=4$.
}
\label{f-fusi}
\end{figure}

%
%
A similar qualitative behavior to the special case \(V=0\), is found when $|V/\alpha_0|<1$. The solutions of 
(\ref{exotic}) show that $X(t)$ converge towards 
a fixed value
$$X_f=(-1)^n\arcsin\left(\frac{V}{\alpha_0}\right)+m\pi\:.$$
The difference here is that the dipole is trapped, and the wave forces the $x$-component of the dipole speed
to the phase velocity $\omega/k$,
 by adjusting the slope of the wave that the dipole will 
``see'', Fig.~\ref{f-dip}(a). On the other hand,
if $|V/\alpha_0|>1$, the dipole has enough strength to ``surpass''
the
 wave with $V$ as the 
averaged $x$-component of its speed, Fig.~\ref{f-dip}(b). The wave has only a small 
influence on the dipole.

%
%

\begin{figure} 
\includegraphics[width=0.45\textwidth]{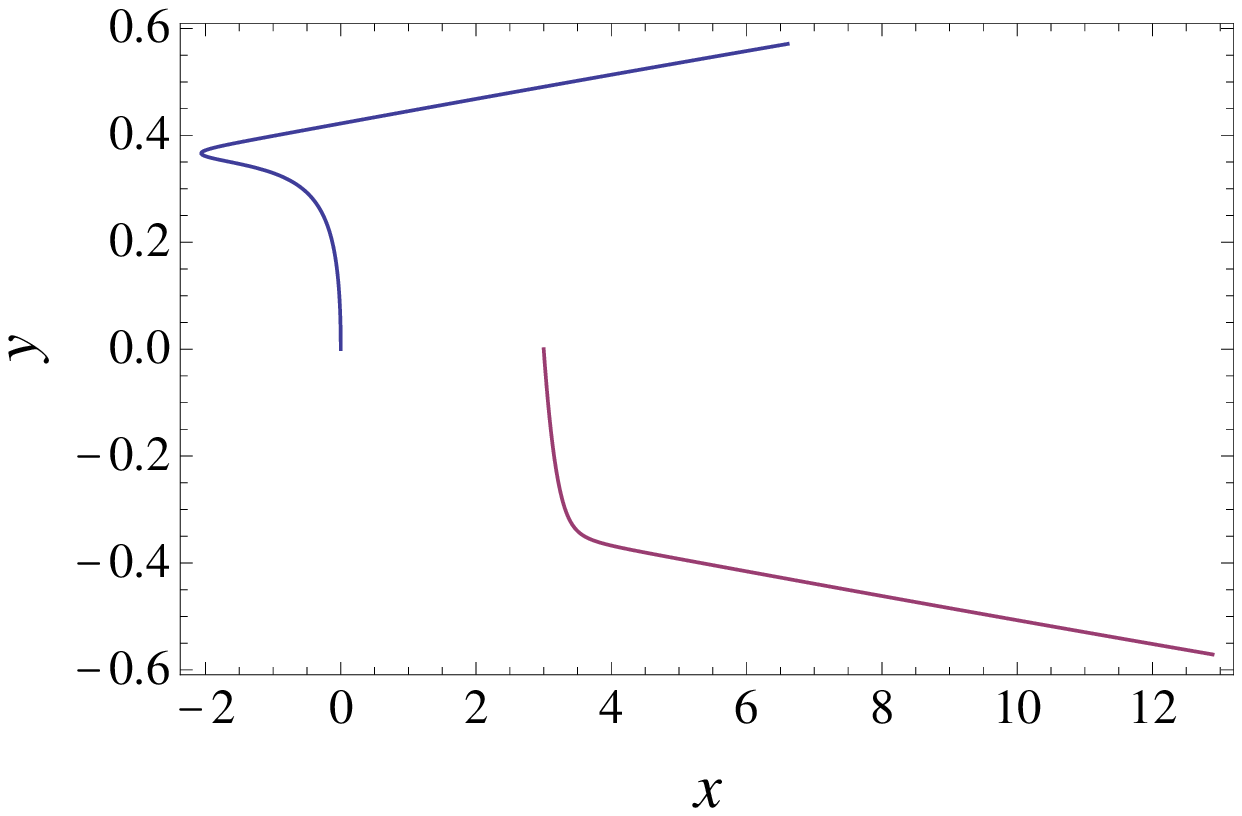}(a)\hfill%
\includegraphics[width=0.45\textwidth]{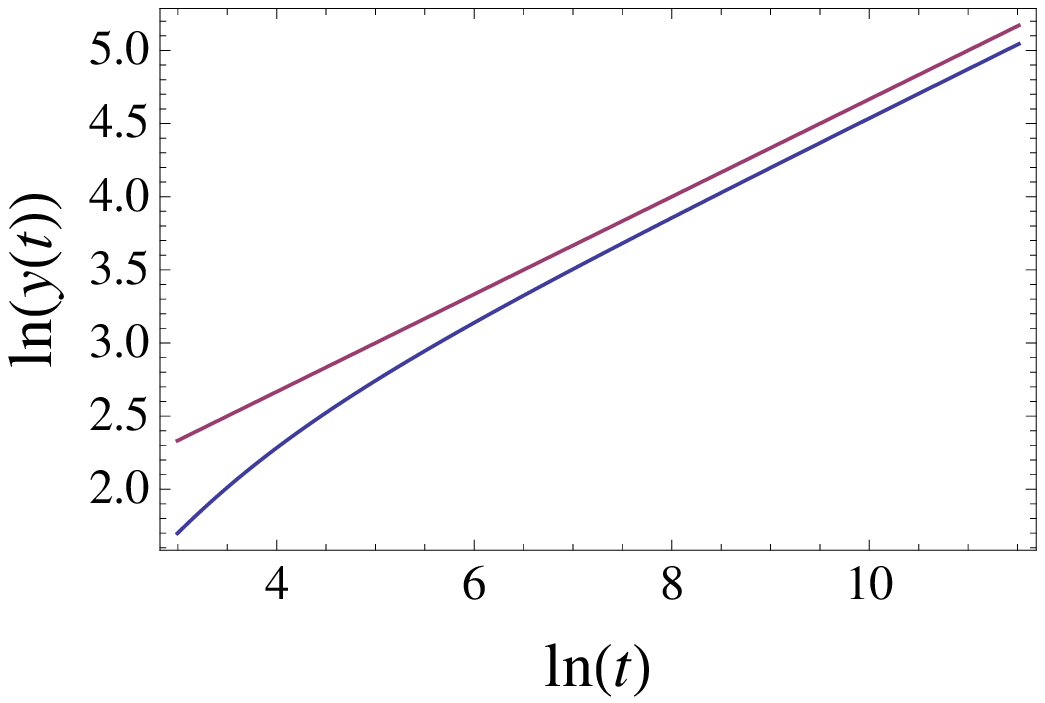}(b)
\caption{(Color online) Trajectory of the monopole leading to the splitting. (a) Typical splitting process when the wave length is smaller than the vortices separation. (b) Long-time behavior, the straight line corresponds to a power-law fit with exponent $1/3$. Parameters are similar to the ones of Fig.~\protect\ref{f-fusi}, but with a wave phase of \(\pi\).
}
\label{f-split}
\end{figure}

In the monopole case it is convenient to
introduce polar coordinates: $x=r\cos\theta$, $y=r\sin\theta$, to obtain the system,
\begin{subequations}
\label{mon} 
\begin{eqnarray}
\label{mon-r}
\dot r&=&-\alpha_0\cos X\sin(r\cos\theta)\cos\theta\,,\\
\label{mon-t} 
r\dot\theta&=&\frac{\alpha_{\!+}}{r}+
       \alpha_0\cos X\sin(r\cos\theta)\sin\theta \,,\\
\label{mon-t1}
\dot X&=&-1-\alpha_0\sin X\cos(r\cos\theta)\,,\\
\label{mon-Y}
\dot Y&=&0\,.
\end{eqnarray}
\end{subequations}
These three autonomous nonlinear equations allow chaotic behavior of
the vortex trajectories, the external wave destroying the vortex
integrals of motion. The variety of possible trajectories is illustrated in the series of Figs.~\ref{f-qper}, \ref{f-fusi}, and \ref{f-split}. We introduce various simplifying assumptions in order to qualitatively describe the behavior of the monopole in these different physical regimes. We 
first investigate the long wavelength limit in both weak and finite wave amplitude approximations, and then we discuss the small 
wave length behavior. We particularly 
focus on the merging or splitting trajectories, in the presence of a finite amplitude wave.

%
%

In the weak long-wave regime, we can assume that 
$|\alpha_0|\sim r\ll 1$, and $|\alpha_{\!+}|\sim 1$; the consistency 
of these assumption will be valid as long as $r$ remain within the 
neighborhood of $r_0$. A series perturbation development gives
then, 
\begin{equation}
r=r_0\,,
\quad
\theta=\omega_0 t\,,
\end{equation}
\begin{equation}
X=t -\alpha_0\cos t\,,
\quad
Y=Y_0\,,
\end{equation}
where $\omega_0=\alpha_{\! +}/{r_0}^2$ is the natural frequency of
the 
unperturbed monopole.
The vortex trajectories are therefore confined in a small region 
of maximum thickness $2\alpha_0$ around the circle.

%
%

In the case where $r\ll 1$ and 
$|\alpha_0|\sim|\alpha_{\!+}|\approx1$ (the wave amplitude is finite, comparable to the vortex intensity), the motion equations become,
\begin{subequations}
\label{monopeq}
\begin{eqnarray}
\dot{\theta}&=&\frac{\alpha_+}{r^2}\,,
	\label{monopT}\\
\dot{r}&=&-\alpha_0 r\cos^2\theta\cos X\approx 
	-\frac{\alpha_0}{2}r \cos X\,,
	\label{monopR}\\
\dot{X}&=&-1-\alpha_0\sin X
	\label{monopX}\,,
\end{eqnarray}
\end{subequations}
where in the distance equation (\ref{monopR}), we approximated the cosine term by its mean value, using the fact that in the limit $r\rightarrow0$, $\theta$ given by (\ref{monopT}) varies rapidly. We then obtain for $X$ an equation similar to (\ref{exotic}) with 
$V=1$; it can be readily integrated to give
\begin{equation}
X(t)=-2 \arctan\left[
	\alpha_0+s\tan\left(\frac{s}{2}(t-t_0)\right)
\right]
\end{equation} 
for $\alpha_0<1$, and for $\alpha_0>1$,
\begin{equation}
X(t)=-2 \arctan\left[
	\alpha_0-s\tanh\left(\frac{s}{2}(t-t_0)\right)
\right]\,,
\end{equation} 
where $s=|\alpha_0^2-1|^{1/2}$. 

%
%

In the case, $\alpha_0<1$ and $r\ll1$, quasi-periodic motion is possible. Indeed, for weak enough wave amplitude, one retrieves the perturbation result, with  $X(t)\approx X_0(t)=t(1-\alpha_0^2)/(1+\alpha_0^2)-2 \arctan(\alpha_0)$ linear and 
\begin{equation}
r\approx r_0\exp\left[-\frac{\alpha_0(1+\alpha_0^2)}{2(1-\alpha_0^2)}\sin X_0(t)\right]\,,
\end{equation}
corresponding to a periodic separation superposed to a constant drift $(\dot{X}(t),0)$ in the \(x\)-direction, as illustrated in Fig.~\ref{f-qper}a. Increasing the initial distance between the vortices, one can enter a region where the trajectories become chaotic (see Fig.~\ref{f-qper}b).

%
%
The other case, $\alpha_0>1$, corresponds to the merging of the two vortices, as can be seen in Fig.~\ref{f-fusi}. The motion of the center of mass  $X$ converges to a fixed value $X_f=-2\arctan(\alpha_0-\sqrt{\alpha_0^2-1})$; as a result, the distance $r$ satisfies the long-time solution,
\begin{equation}
r\rightarrow r_0\exp\left[-\frac{\alpha_0}{2}\cos(X_f)t\right]\,,
\end{equation}
where $\cos X_f$ is positive for $\alpha_0>1$. Therefore, we obtain in this case an exponential approach of the two vortices, accompanied with a diverging angular velocity $\dot\theta\rightarrow\infty$ (Fig.~\ref{f-fusi}b). The merging characteristic time is $2/\alpha_0 \cos X_f(\alpha_0)$, and within the present approximation, it diverges when $\alpha_0\rightarrow1$. This is consistent with the vortex merging observed in the numerical solutions, as seen in Fig.~\ref{f-fusi}.

%
%

Let us now consider the situation where, $r\gg 1$ 
(the wave length is small compared to the size of the monopole), 
$1\lesssim\alpha_0\sim\alpha_{\!+}\ll r$, such that the first term in (\ref{mon-t}) may be neglected. This case corresponds to the vortex splitting process, shown in Fig.~\ref{f-split}. Equations (\ref{mon}) become 
now, $\theta\approx\mathrm{const}$ and  
\begin{subequations}
\label{split-eq}
\begin{eqnarray}
\label{split-x}
\dot{x}&=&-\alpha_0\cos X\sin x\\
\label{split-X}
\dot{X}&=&-1-\alpha_0\sin X\cos x\\
\label{split-y}
\dot{y}&=&\frac{\alpha_+ x}{x^2+y^2}\\
\end{eqnarray}
\end{subequations}
similar to Eqs.~(\ref{dip}) for the dipole, but with a difference in the equation for $y(t)$. Asymptotically (\ref{split-x},\ref{split-X}) give $x(t)+X(t)=\mathrm{const.}$, which justifies the approximation
\begin{equation}
\dot{y}\approx\frac{\mathrm{const}}{y^2},\quad y(t)\sim t^{1/3}
\label{split-yt}
\end{equation} 
Indeed, if $r$ is initially large, Eq.~(\ref{split-yt}) shows that it remains large at long times. We confirmed the validity and verified by direct numerical integration of the vortex equations (\ref{num}), that the asymptotic behavior of the vortex distance follows a power-law with the exponent $1/3$ (see Fig.~\ref{f-split}b).

\begin{figure} 
\centering
\includegraphics[width=0.30\textwidth]{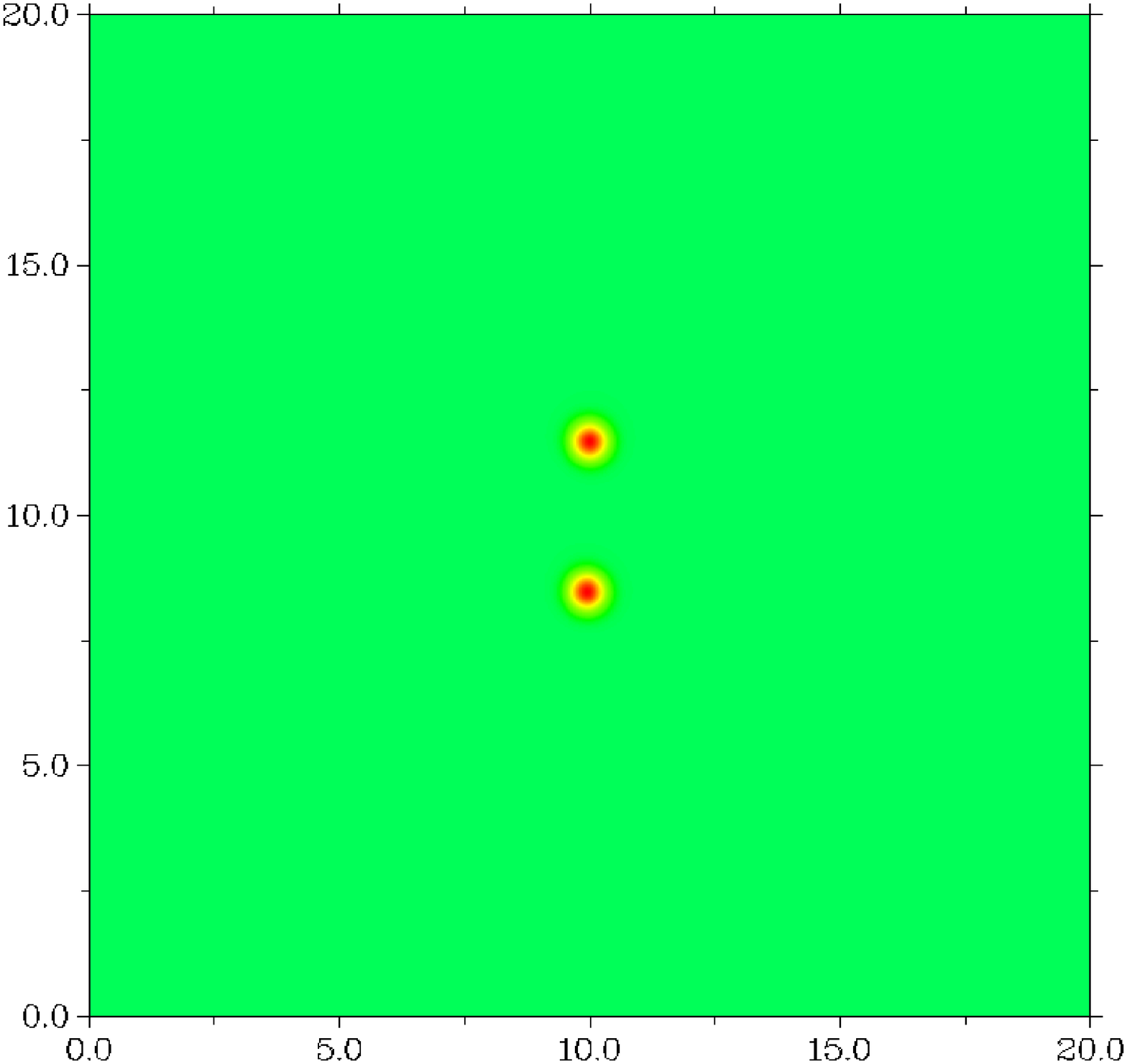}\hfill%
\includegraphics[width=0.30\textwidth]{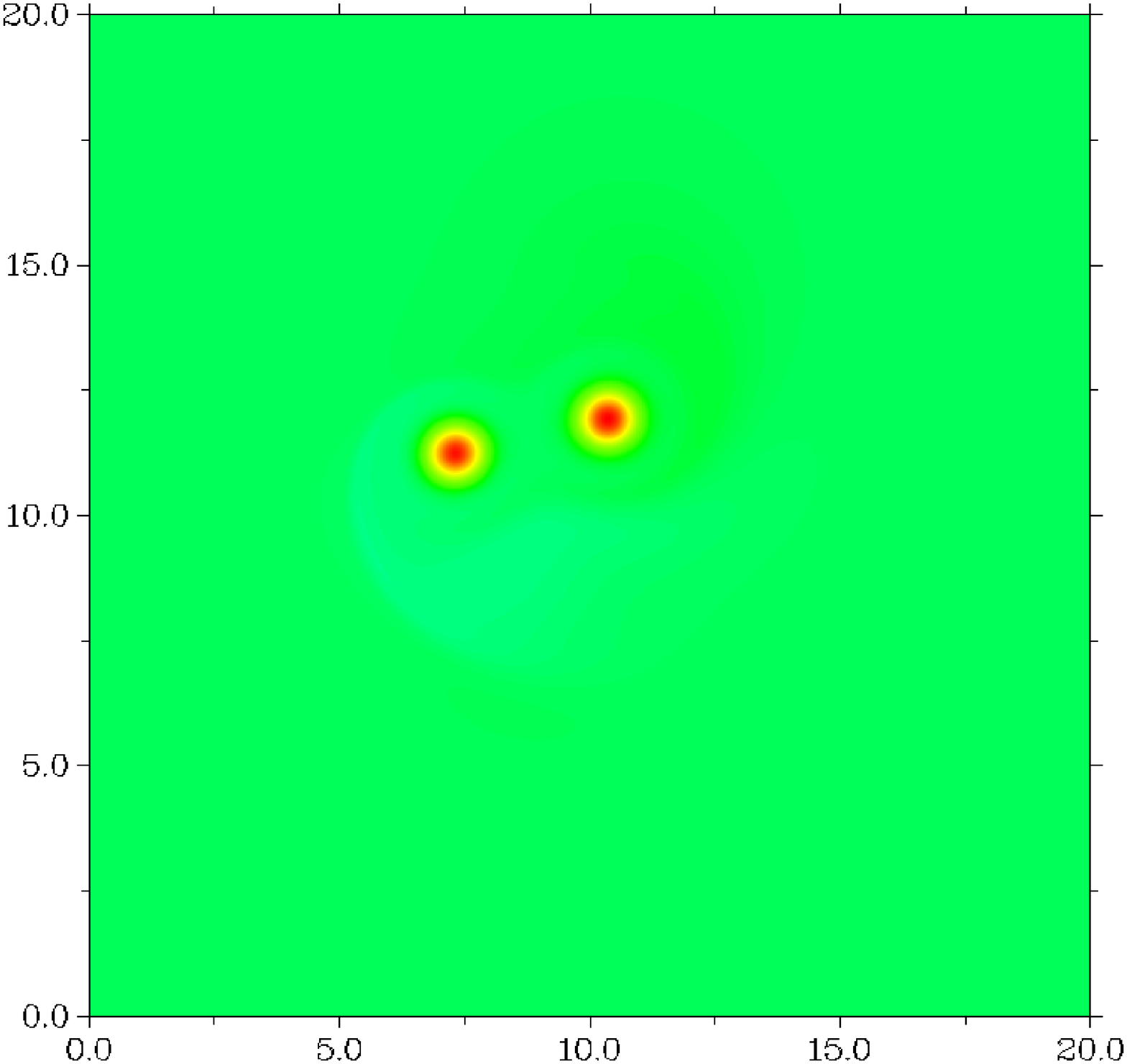}\hfill%
\includegraphics[width=0.30\textwidth]{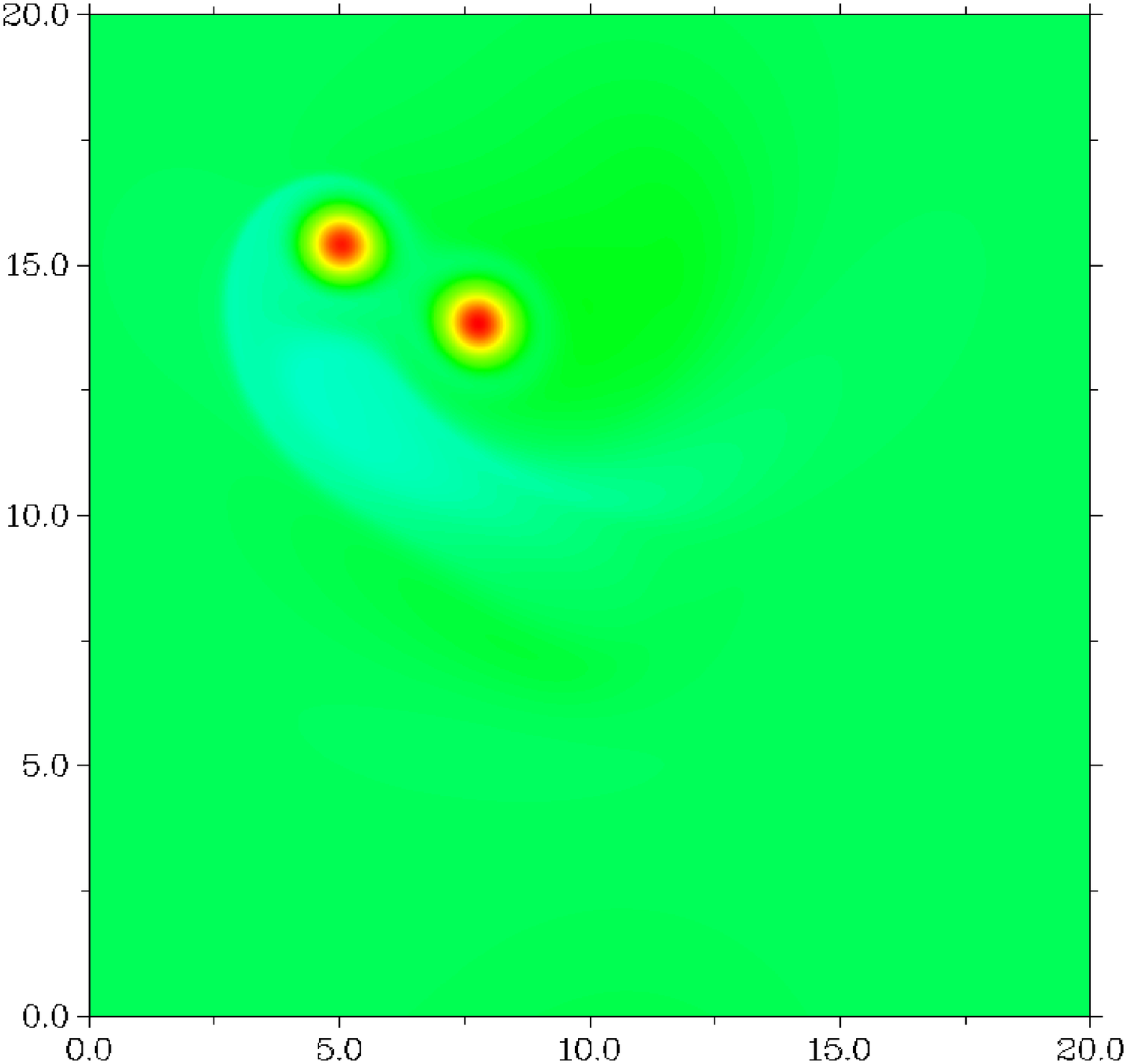}
\caption{(Color online) Potential vorticity of two gaussian monopoles at times 0 (left), 7.5 (center) and 15 (right), in units of \(\rho_s\) and \(v_d\). Without the interaction with a wave, the vortices rotate and drift; the small diffusion of the vorticity is due to the presence of a viscosity term in (\ref{CHM}).}
\label{f-free}
\end{figure}
\section{Wave-vortex interaction in the Charney-Hasegawa-Mima dynamics}
\label{Sec:HM}

\begin{figure*} 
\centering
\includegraphics[width=0.11\textwidth]{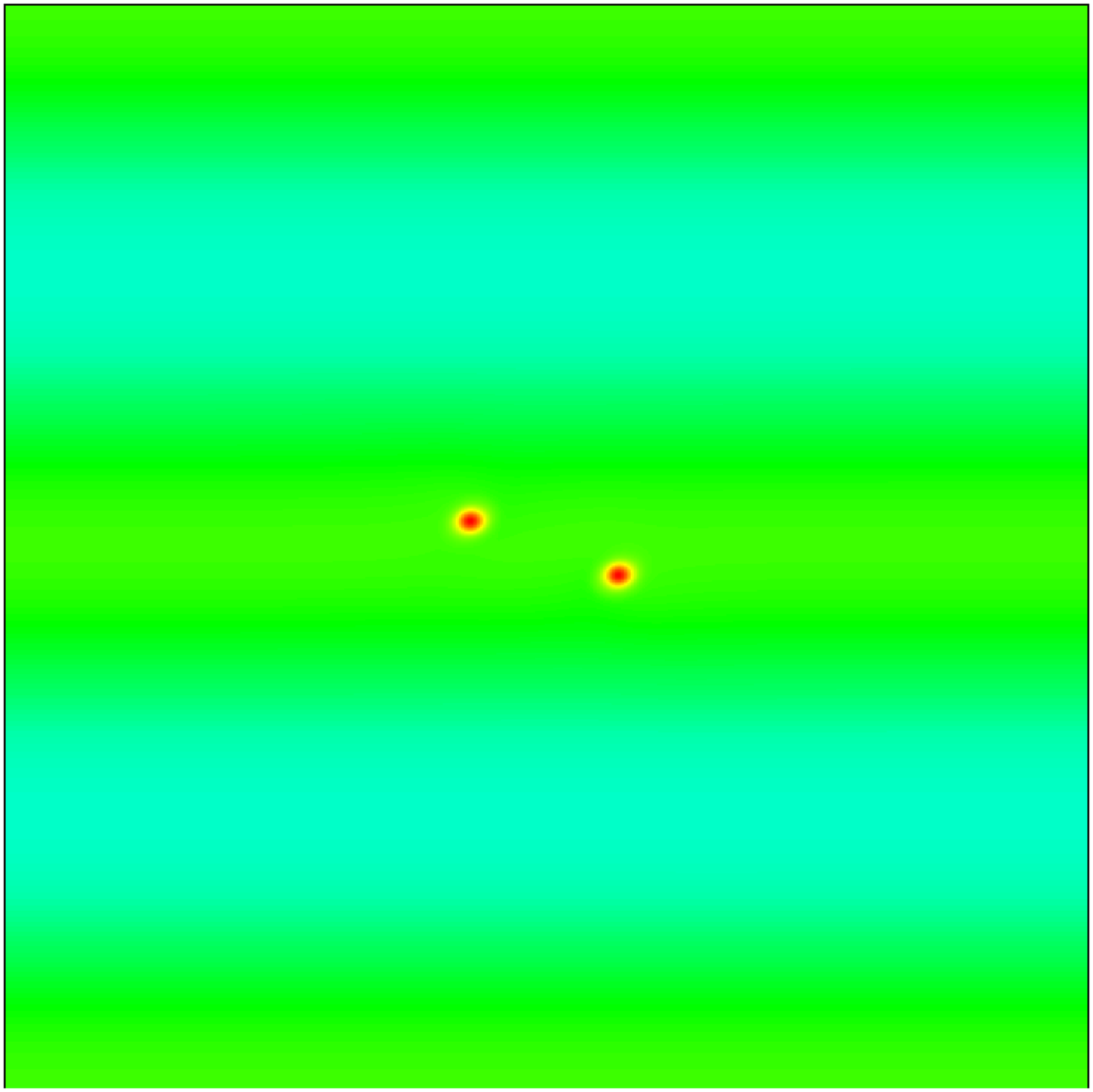}%
\includegraphics[width=0.11\textwidth]{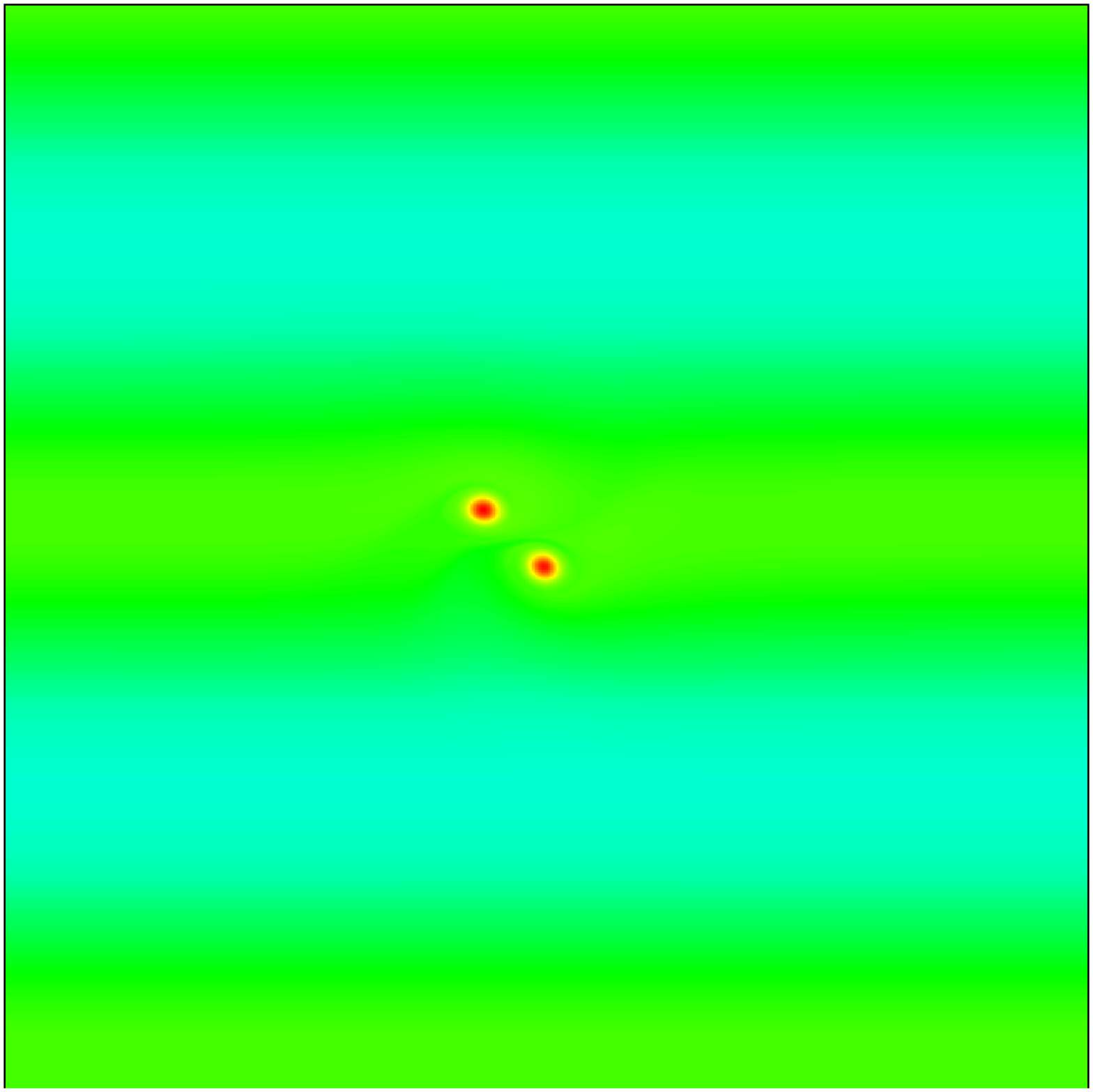}%
\includegraphics[width=0.11\textwidth]{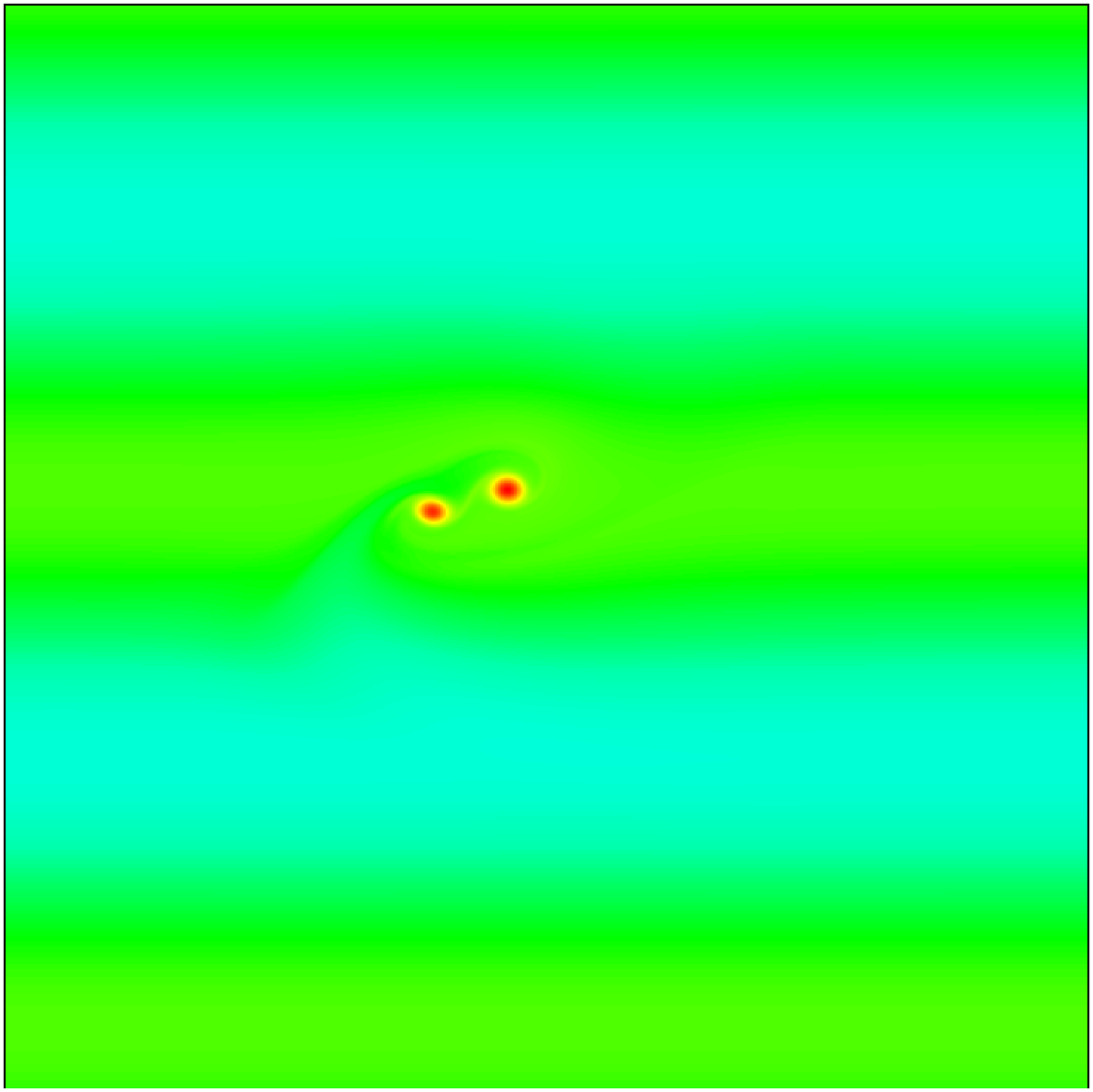}%
\includegraphics[width=0.11\textwidth]{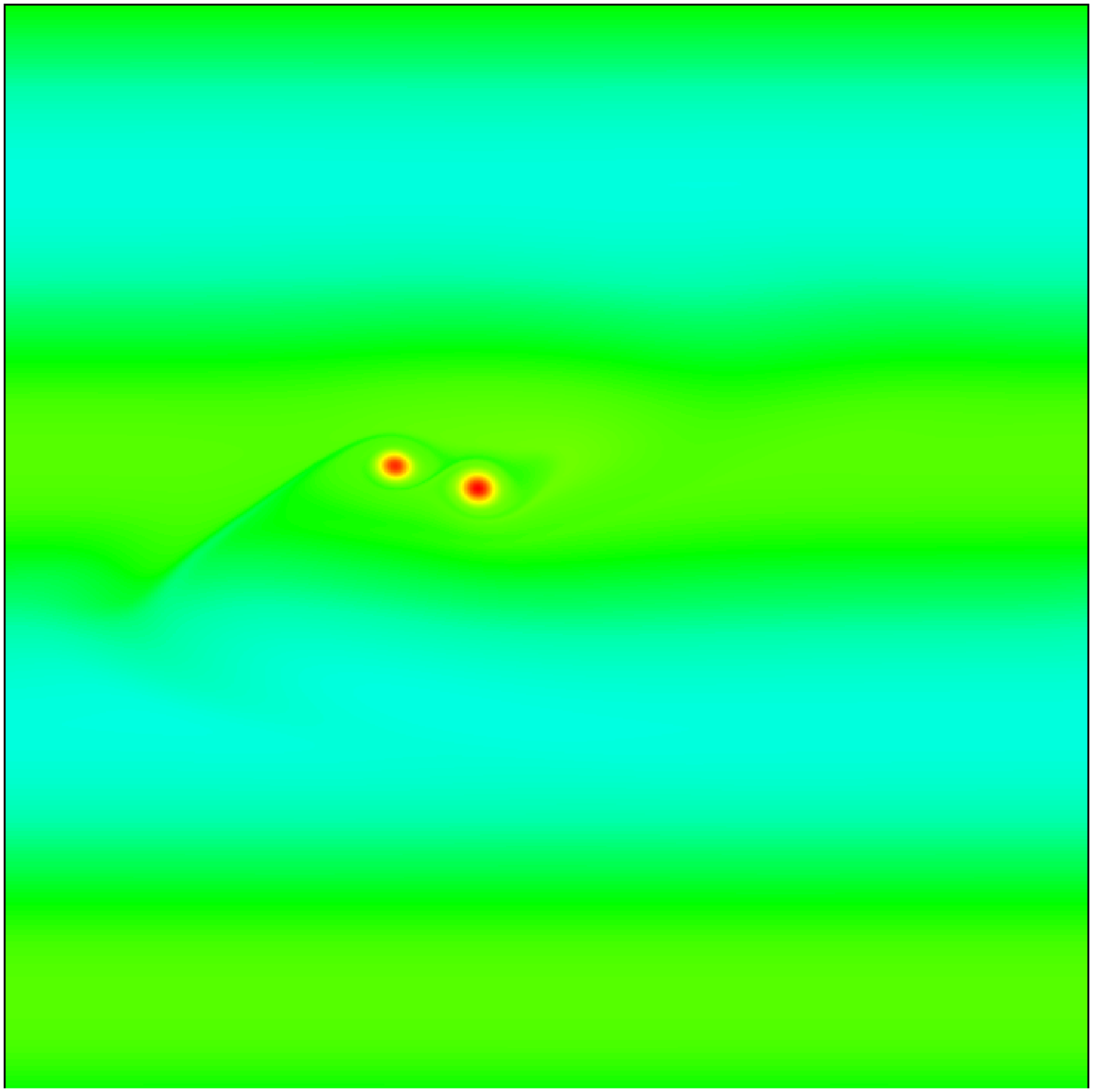}%
\includegraphics[width=0.11\textwidth]{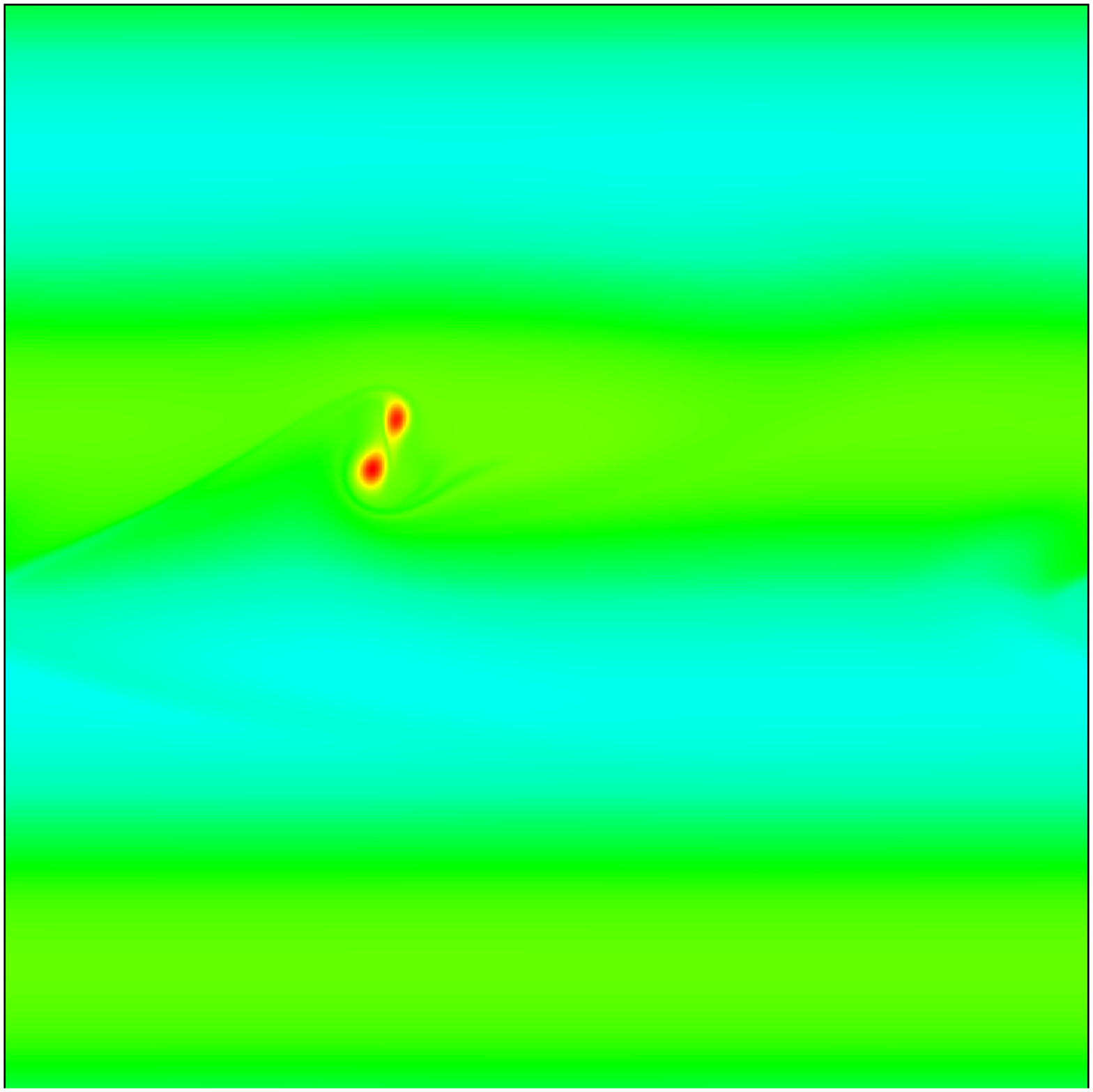}%
\includegraphics[width=0.11\textwidth]{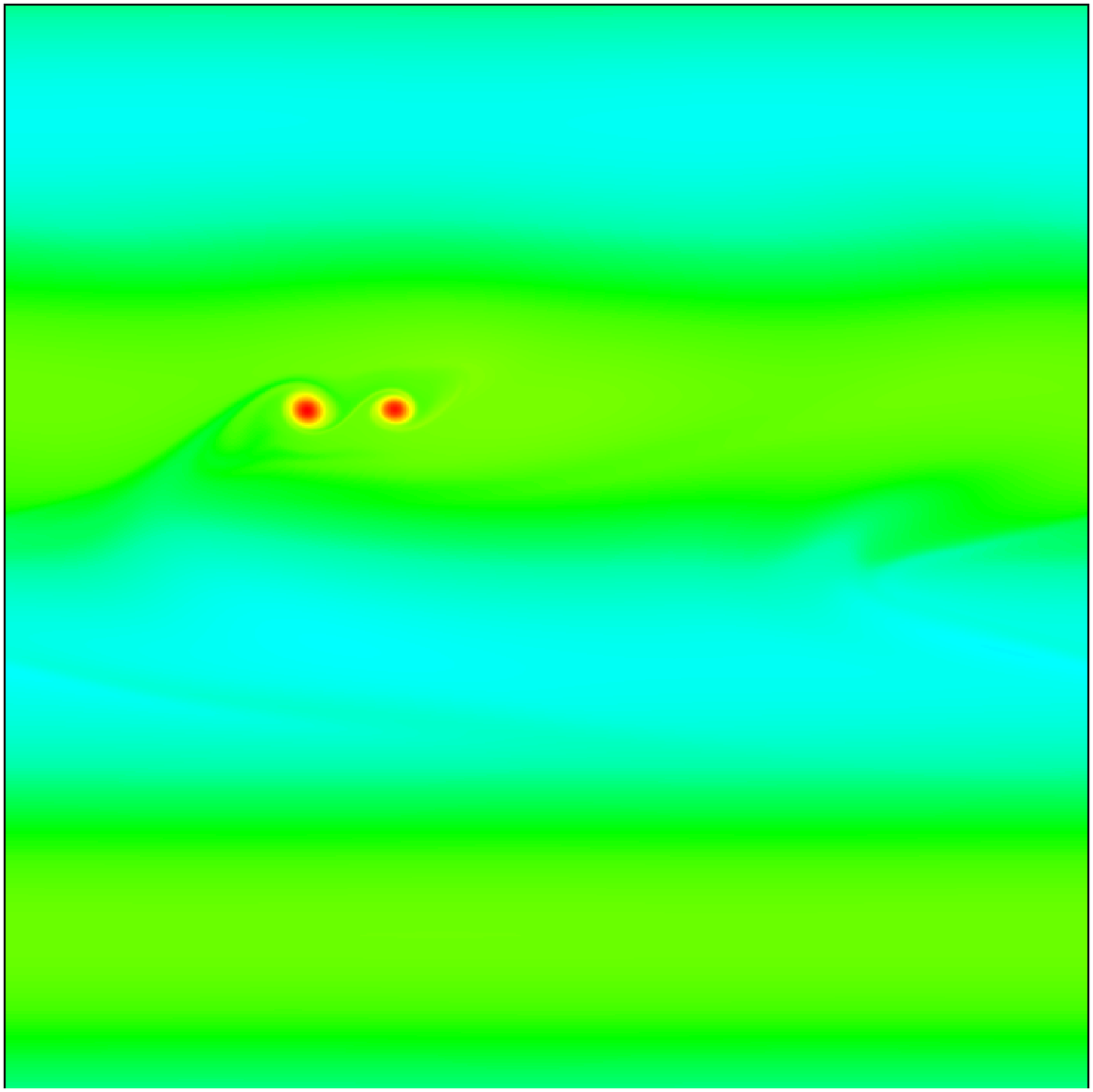}%
\includegraphics[width=0.11\textwidth]{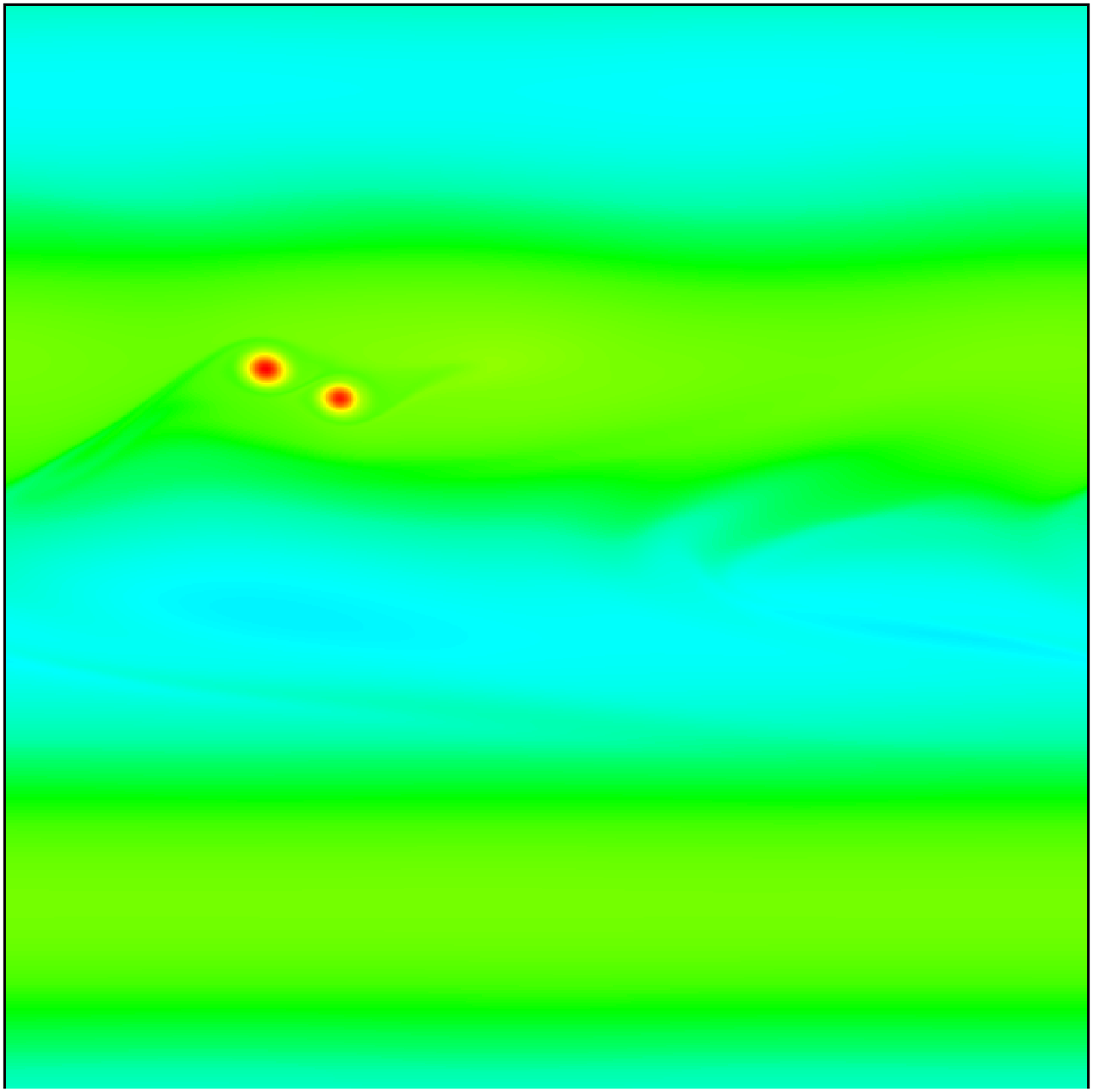}%
\includegraphics[width=0.11\textwidth]{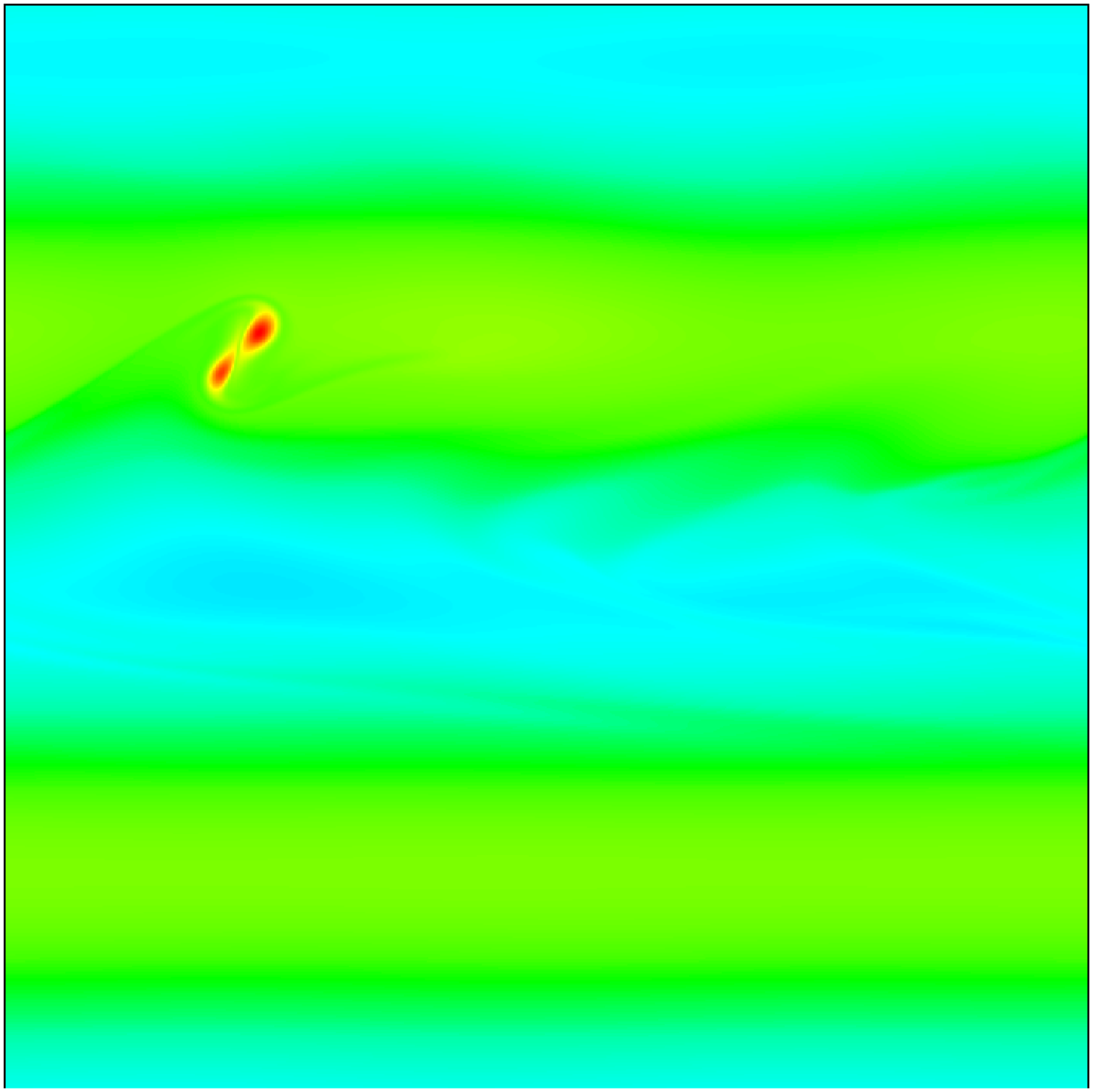}%
\includegraphics[width=0.11\textwidth]{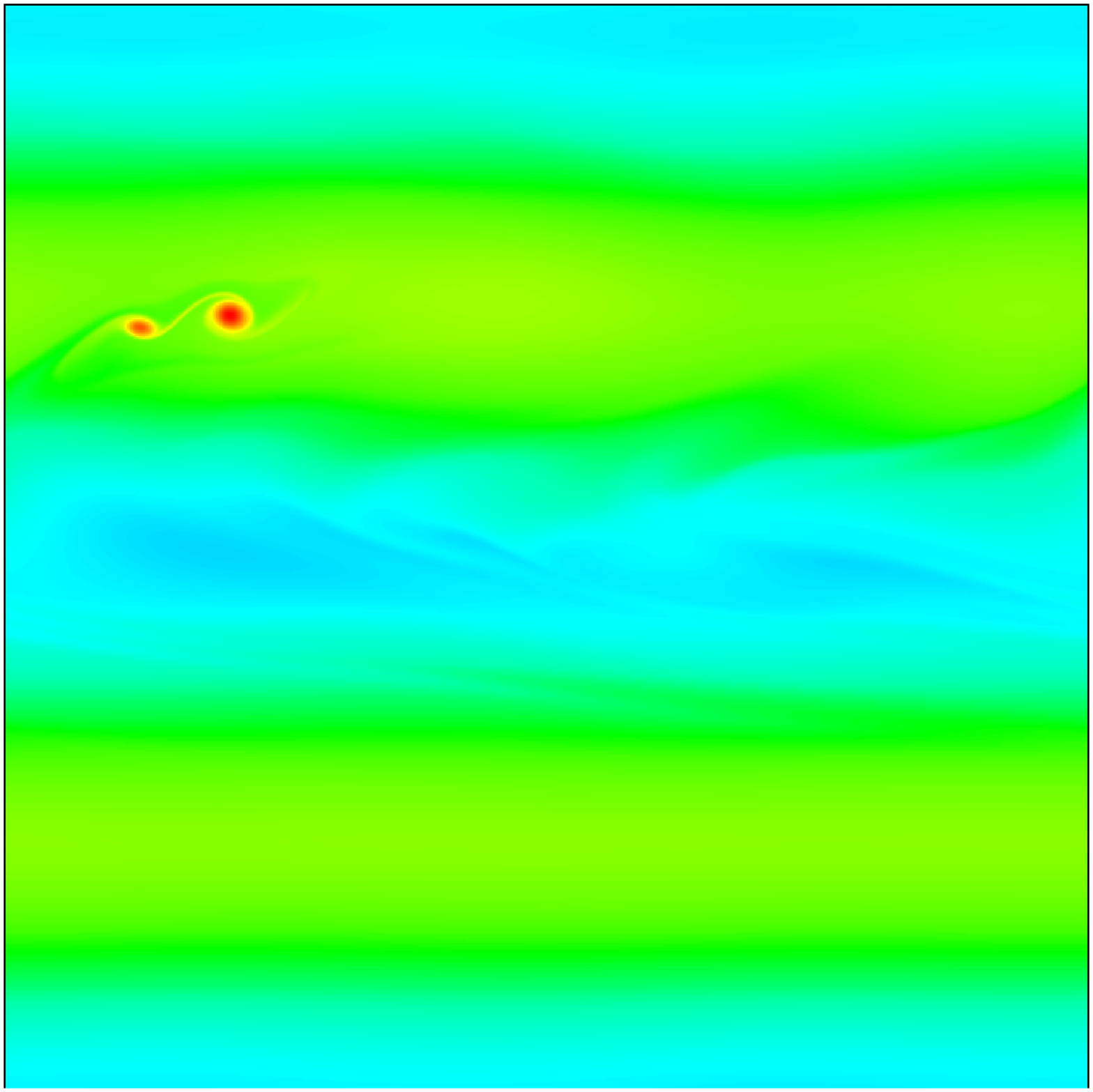}
\caption{(Color online) Potential vorticity of two vortices in the initial field of a transverse wave. The sequence shows the monopole complex trajectory, in the chaotic regime of the corresponding point vortex model. From left to right, \(t=0,\, 0.8,\, 1.6,\, 2.4,\, 3.2,\, 4.0,\, 4.8,\, 5.6,\, 6.4\), square of size \(20^2\), and vortex size parameter \(b=16\).}
\label{f-chaos}
\end{figure*}
%

\begin{figure*} 
\centering
\includegraphics[width=0.30\textwidth]{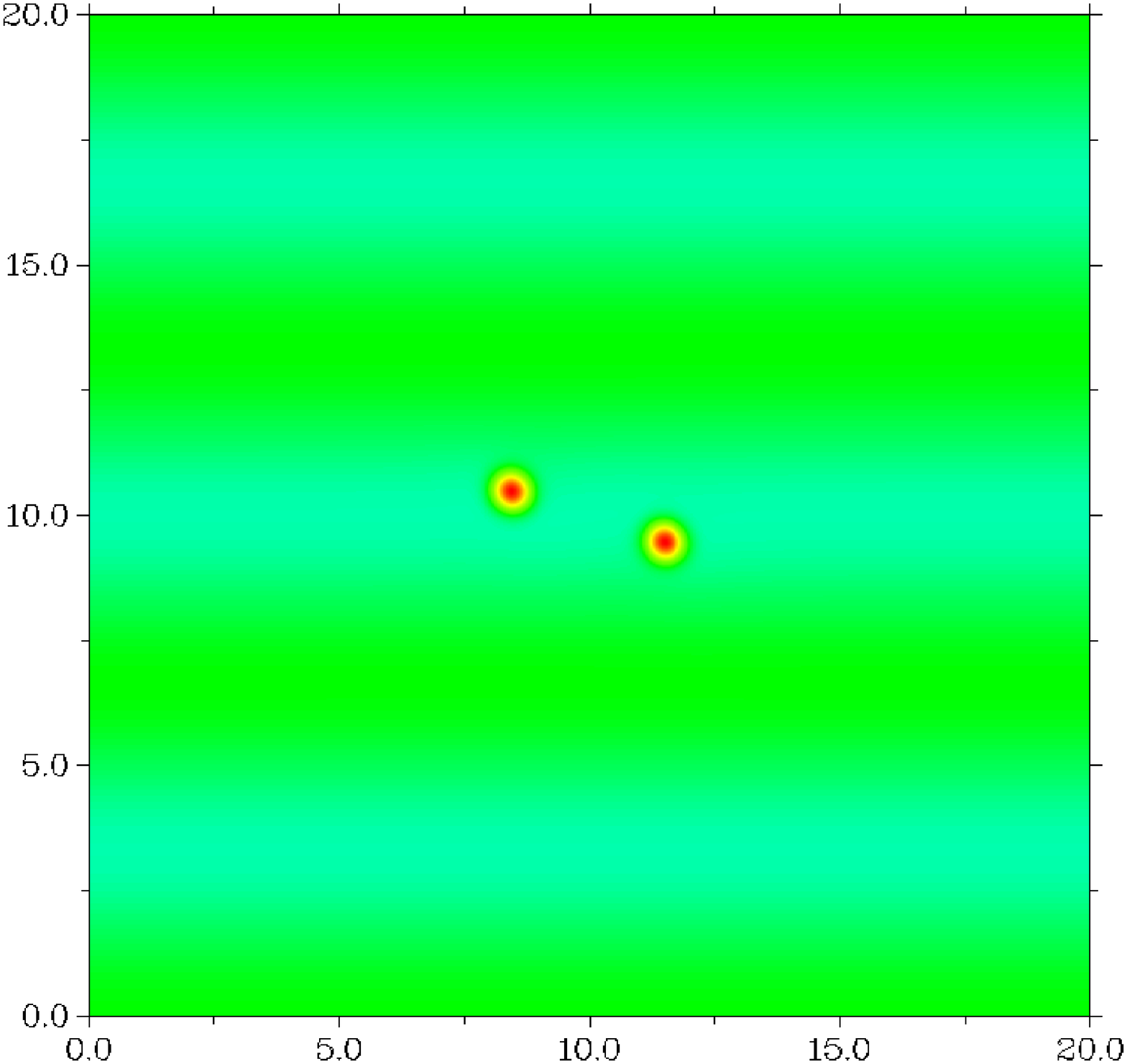}(a)\hfill%
\includegraphics[width=0.30\textwidth]{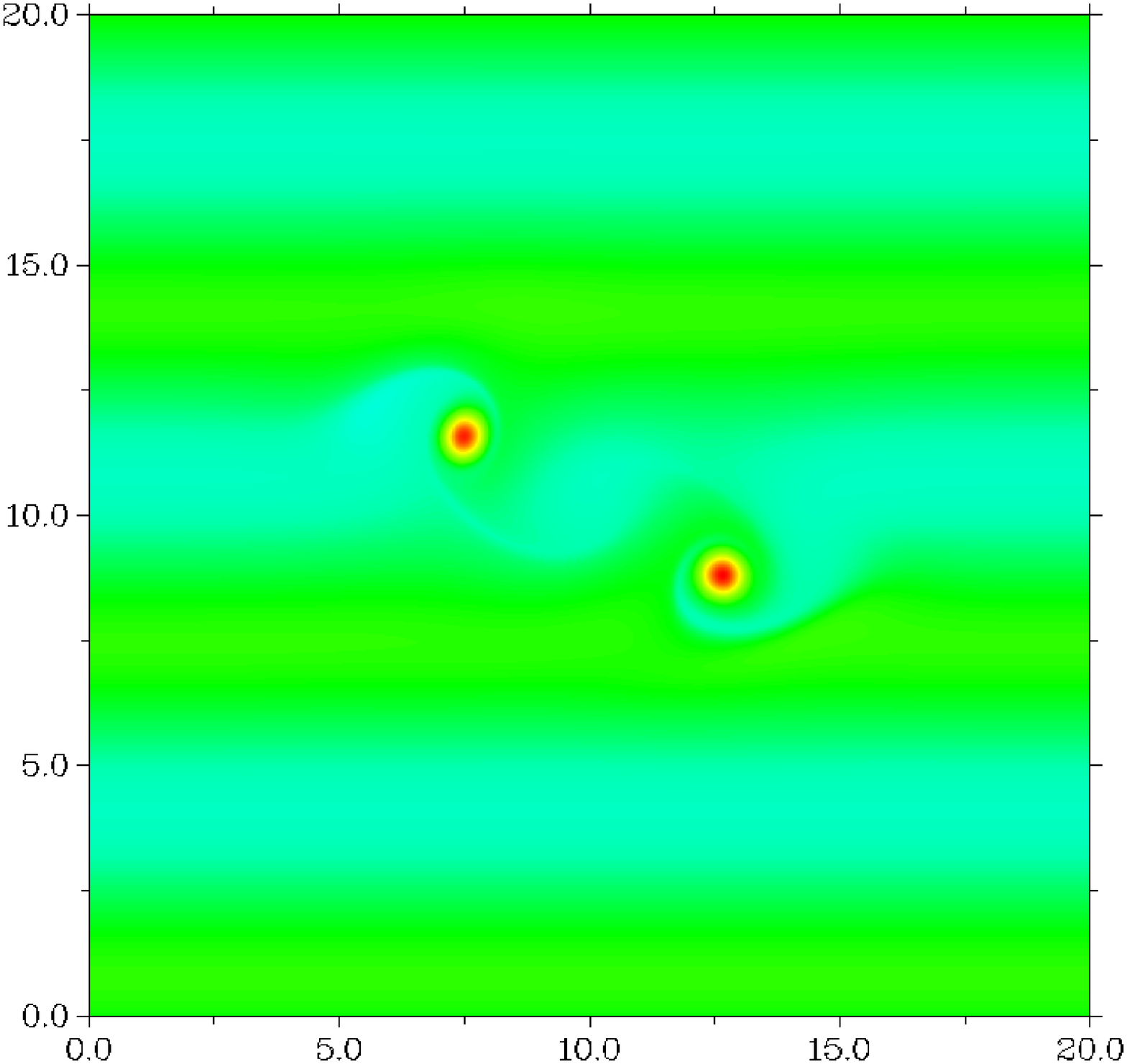}(b)\hfill%
\includegraphics[width=0.30\textwidth]{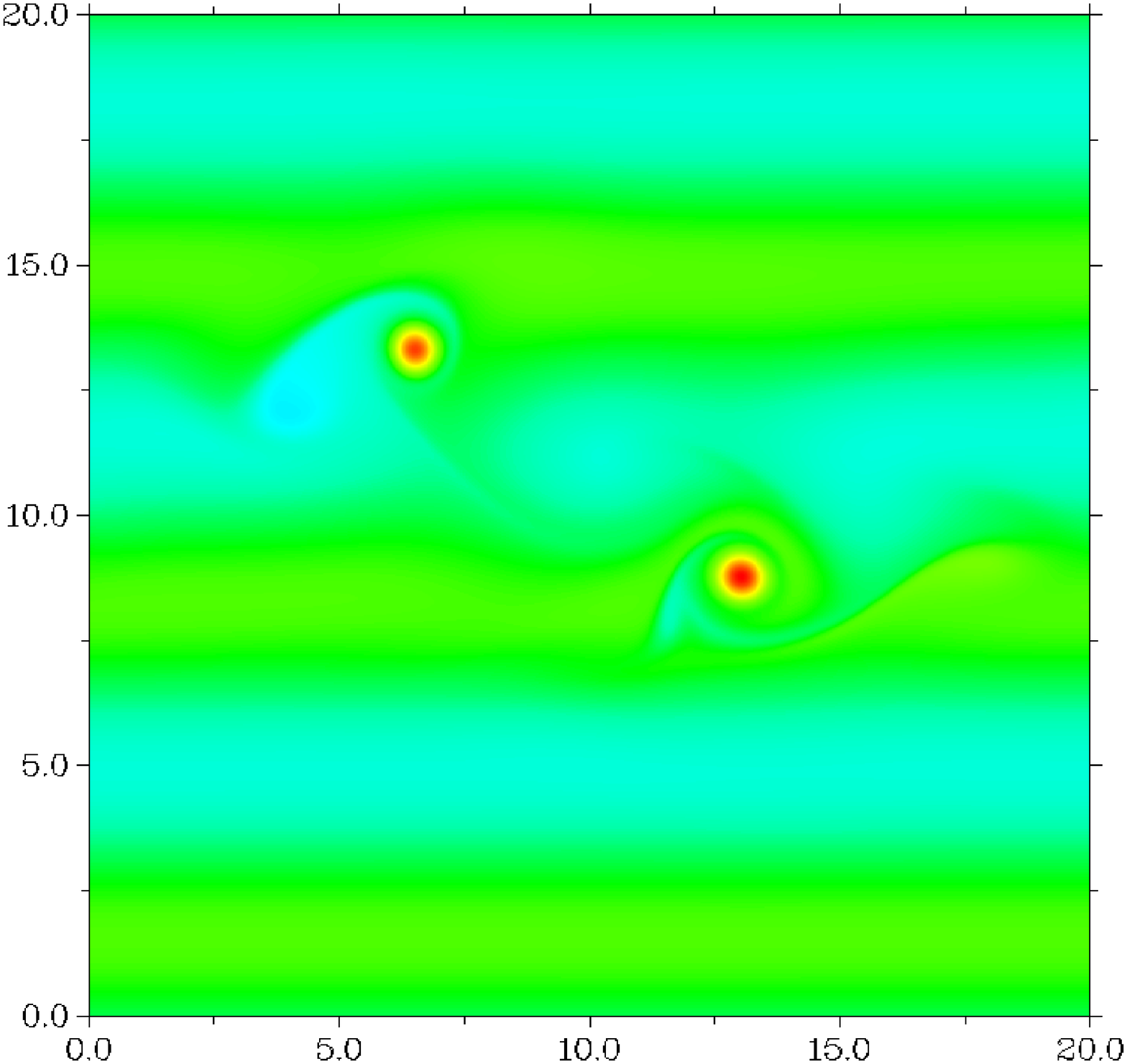}(c)\\
\includegraphics[width=0.30\textwidth]{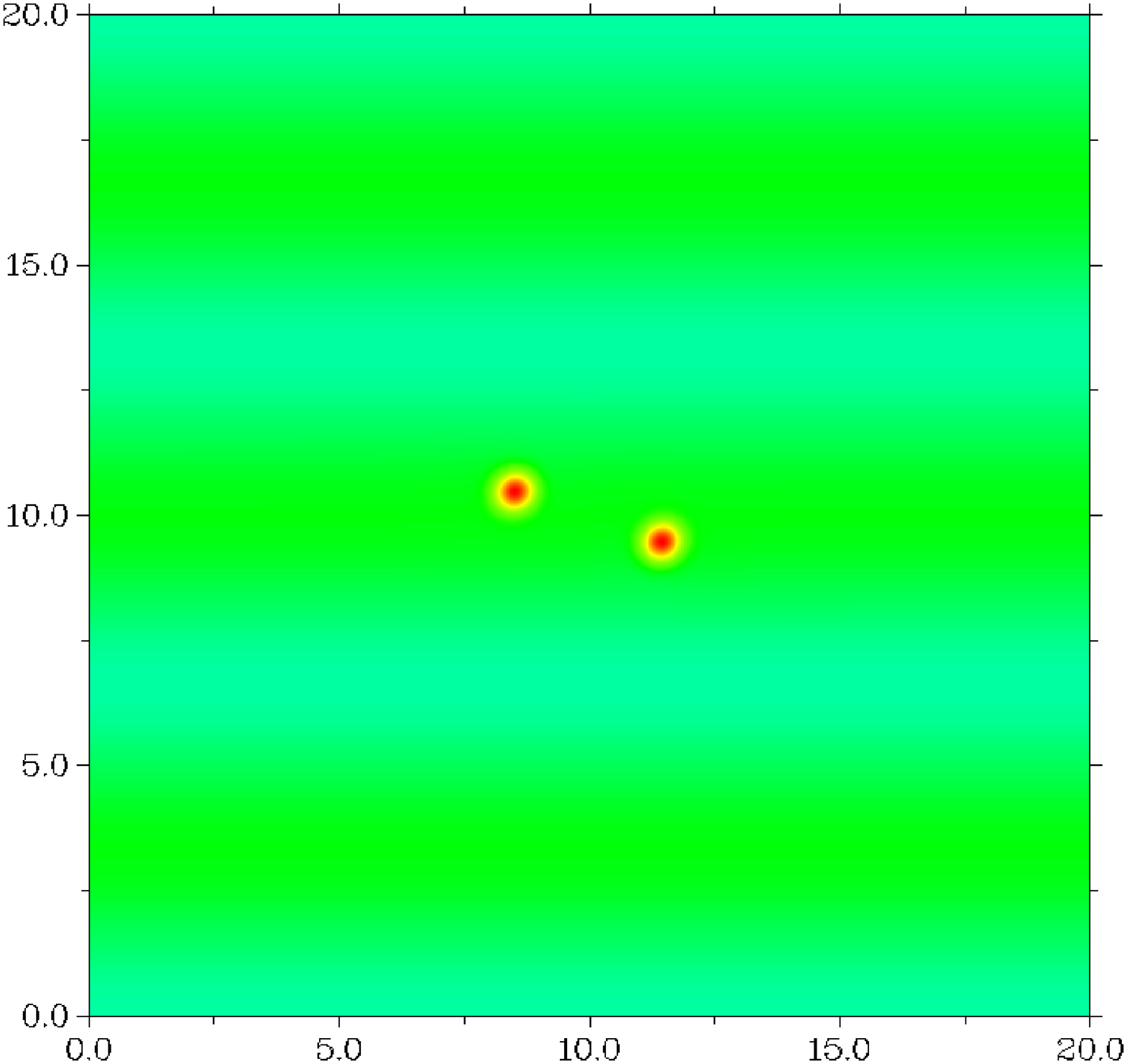}(d)\hfill%
\includegraphics[width=0.30\textwidth]{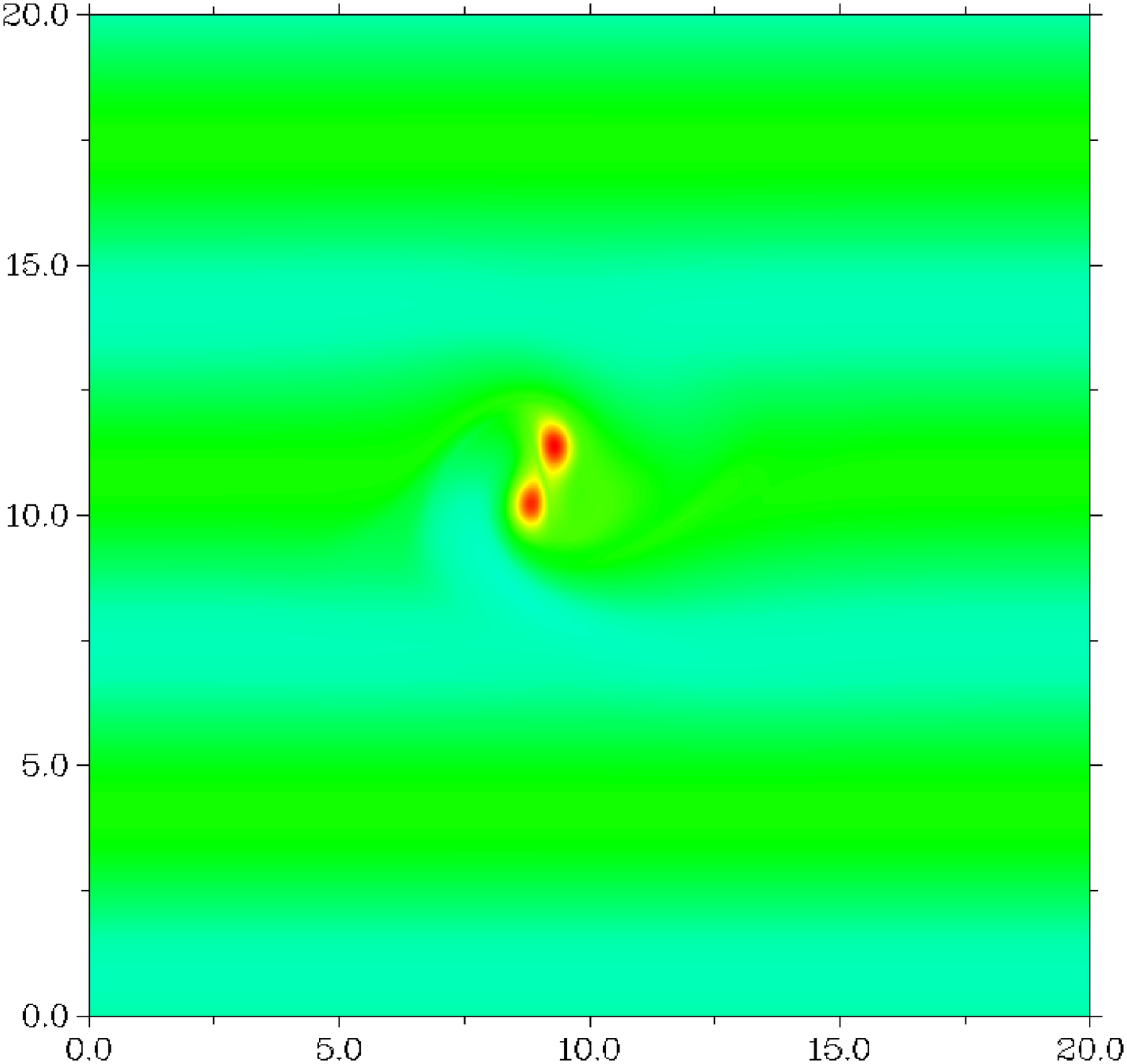}(e)\hfill%
\includegraphics[width=0.30\textwidth]{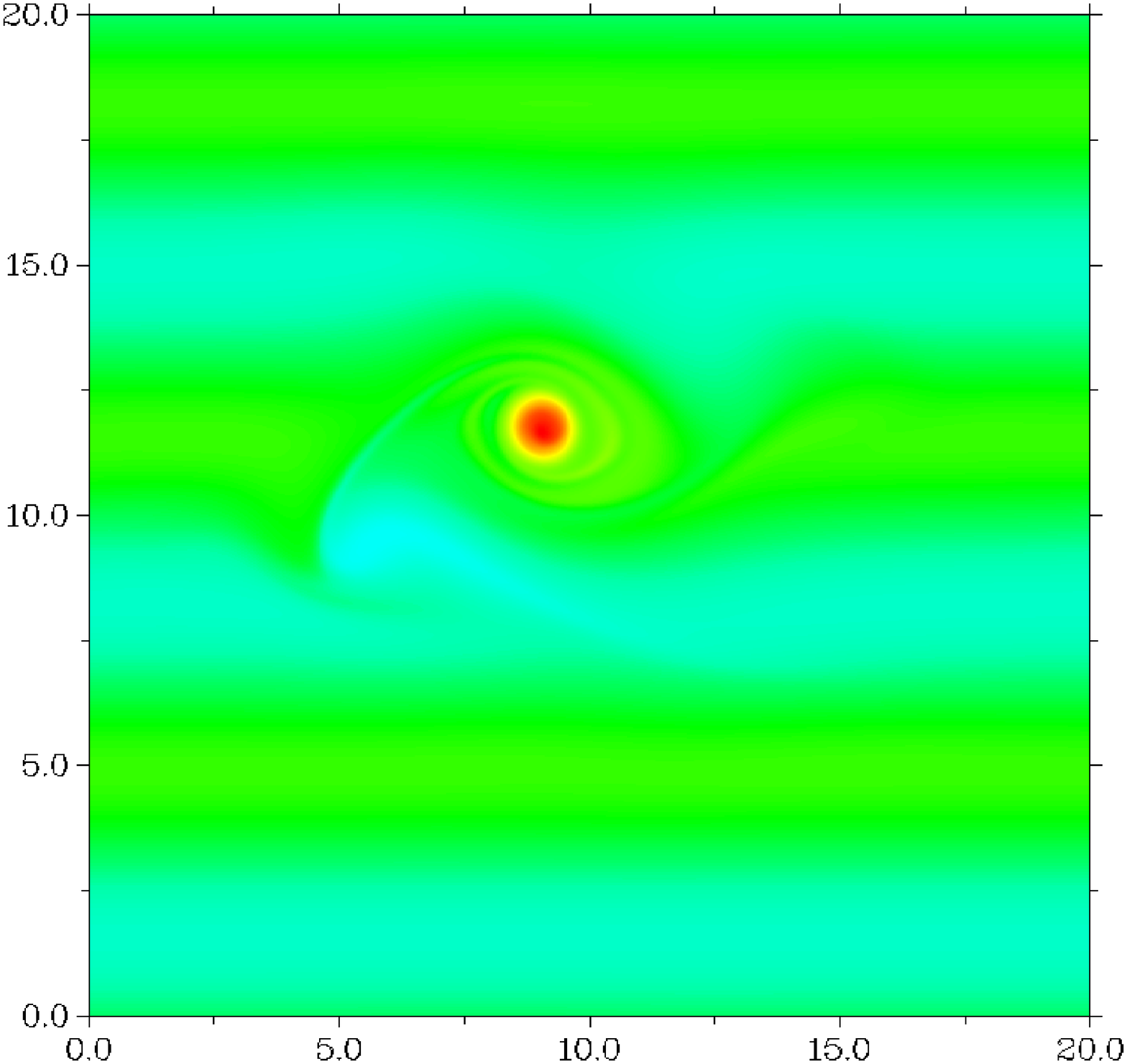}(f)
\caption{(Color online) Evolution of two gaussian monopoles in the transverse wave case. Vortex and wave parameters are as in Fig,~\protect\ref{f-poincare1}. The initial conditions, in (a) and (d), differ only in the phase of the wave with respect to the vortex positions. The first row (a-c) shows the splitting; the second row (d-f) the merging of the vortices.}
\label{f-t}
\end{figure*}

We consider now the Charney-Hasegawa-Mima equation in its two forms, (\ref{CHM}) and (\ref{CHMs}), to study the evolution of two vortices in the presence of a transverse and a longitudinal wave, respectively.
We demonstrated, in the context of the point vortices model, that the presence of a wave can induce a variety of behaviors, including merging, splitting and chaos. However, in real flows or plasma systems, a vortex has a spatial extension and waves have to obey a specific dispersion relation (\ref{dispersion}). Moreover, as mentioned before, the separation of waves and vortices is somewhat arbitrary, their interaction modifies both in such a way that the distinction can be made only within some length scale and time interval. It would then be interesting to analyze how the presence of a wave might influence processes such vortex merging, in the more general framework of the Charney-Hasegawa-Mima equation.

For this purpose we integrate numerically equations (\ref{CHM}) and (\ref{CHMs}),  and we kept the dissipation term to ensure numerical stability. We consider two identical vortices; for each vortex we assign at the initial time, a local vorticity $\Omega_l$ in the form of a Gaussian blob,
\begin{equation}
\Omega_l=-\nabla^2\psi(\bm r,0)+\psi(\bm r,0)/\rho_s^2=\frac{b\Gamma}{\pi}\exp(-br^{2})\:, 
\label{vort_dist}
\end{equation}
($\psi(\bm r,0)$ is the initial stream function), where $b$ controls the extension of the vortex and $\Gamma$ its circulation. In the case of a system governed by (\ref{CHM}), since transverse waves are naturally generated, we shall not impose an external one, but instead add a wave form to the initial vorticity distribution and track its influence on the monopole trajectories. In the case of longitudinal waves, we solve (\ref{CHMs}) with a source term of the form,
\begin{equation}
S(\bm x, t)= S_0\cos(ky-\omega t)\,,
\label{Sw}
\end{equation}
where \(S_0\) measures the source strength (it has dimensions of a square frequency).

Simulations are made using periodic boundary conditions on a square domain of length $L=20$, and the number of spectral modes is typically $1024^{2}$. The numerical code is based on a pseudo-spectral scheme with time stepping a fourth order Runge-Kutta algorithm; the time step is $\delta t=10^{-3}$, and we used a the numerical viscosity of $\mu=0.005$. We measure lengths and velocities in units of \(\rho_s\) and \(v_d\), respectively. The initial state consists on two equal circulation vortices with vorticity distribution (\ref{vort_dist}), where \(b = 8\) and \(\Gamma = 8\pi\). The set of parameters and initial vortex positions, used in the various simulations, correspond to the different cases studied with the point vortex model.

The reference case is displayed in Fig.~\ref{f-free}, where we represent the potential vorticity \(\Omega_l\) of Eq.~(\ref{omega_Lamb}) for three different times, of two monopoles in the absence of initial wave or external source. Without extra perturbations, the two monopoles simply rotate around each other and slowly drift in the \(y\)-direction; the distance between the vortices remains sensibly constant, their shape evolving in the viscous time scale. We note that the observed drift motion is in fact influenced by the emission of a wake, showing that, although Gaussian monopoles are long lived structures, they are not exact solutions of the Charney-Hasegawa-Mima equation.

\subsection{Transverse wave}

Let us first investigate the transverse wave case, where in addition to the two monopoles we superpose an initial wave \(k^2\Gamma_0\cos(ky)\). As it could be inferred from the large chaotic regions observed in the point vortex model (Fig.~\ref{f-poincare1}), for a range of initial conditions the behavior of the two monopoles is unpredictable. In the sequence of Fig.~\ref{f-chaos}, the vortices rotate, approach each other, interchange vorticity, and separate in the background of the wave vorticity. In this case, the initial wave is in phase with the vortices, and as a result, they drift with the wave and rotate at a high frequency compared to the free case of Fig.~\ref{f-free}. The horizontal drift is easily explained by the model (\ref{tnum}) that implies, 
\begin{align}
\label{e-cm}
\frac{d}{dt}X&=\Gamma_0k\sin \omega t\,\cos ky,\nonumber\\
\frac{d}{dt}Y&=0
\end{align}
for the center of mass motion (we use the notation of (\ref{relative})).

The presence of a transverse wave can also trigger processes such as splitting and merging of vortices, as in the point vortex model. In Fig.~\ref{f-t} we present numerical simulations differing in the phase of the initial wave, showing the sensitivity on the initial condition typical of a chaotic system, and leading to split or fusion. We observe that the vortex evolution is much faster than the wave one, as can be asserted by comparing the fusion time of \(t_f\approx 3\) in Fig.~\ref{f-t}f, to the slow rotation of the two free vortices in Fig.~\ref{f-free}.

\begin{figure} 
\centering
\includegraphics[width=0.45\textwidth]{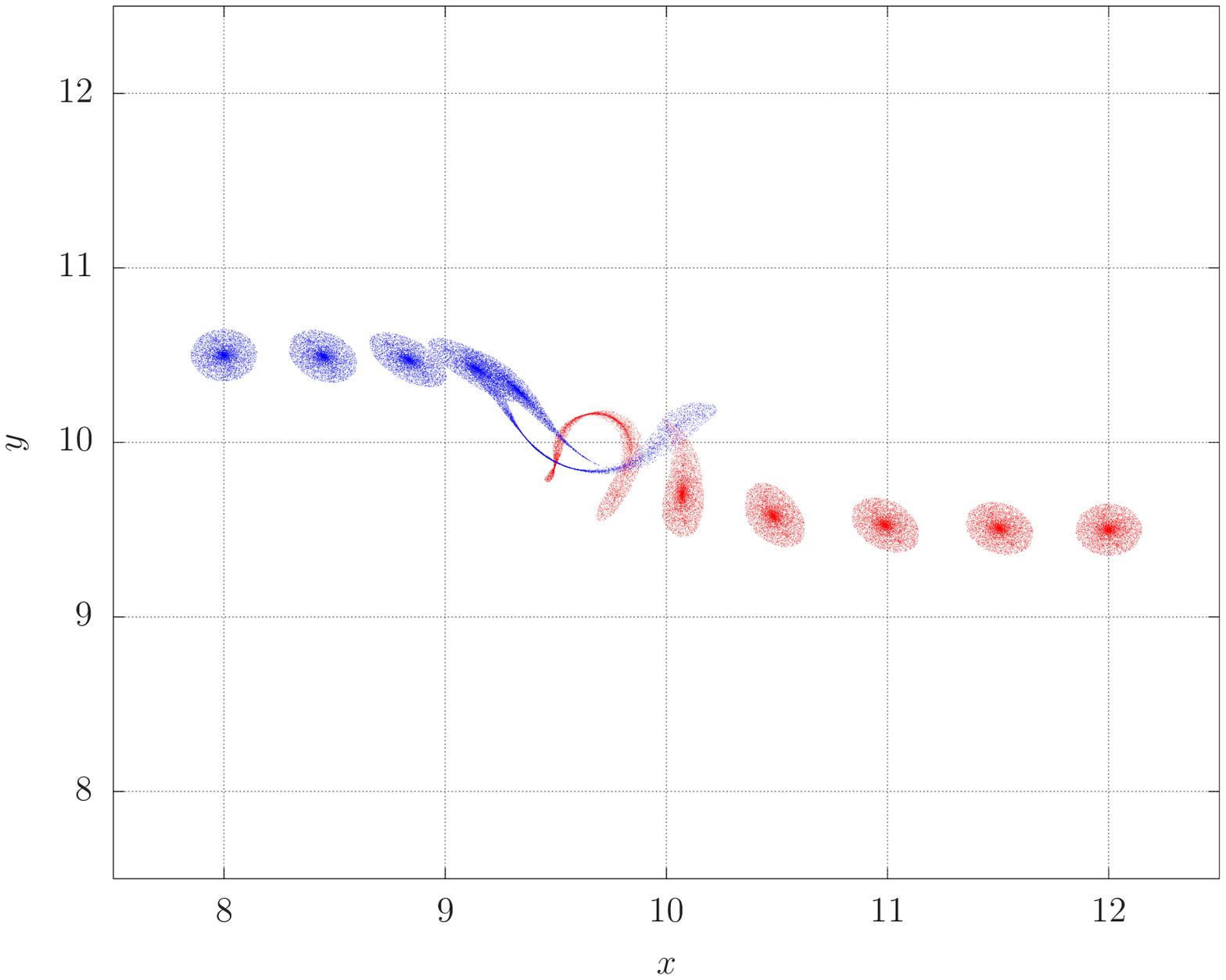}(a)\hfill%
\includegraphics[width=0.45\textwidth]{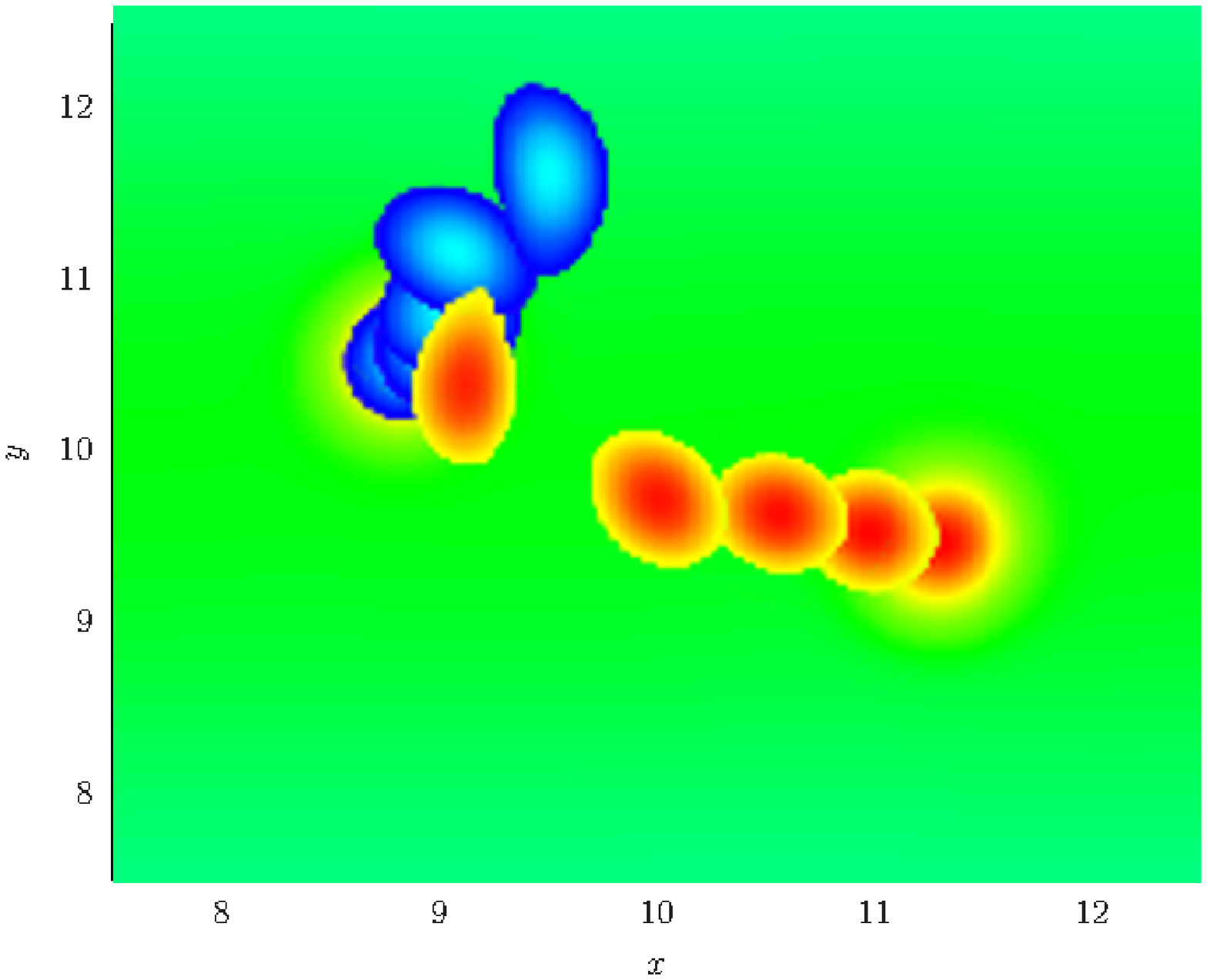}(b)
\caption{(Color online) Comparison of the dynamics of a patch of point vortices (a) with the evolution of two Gaussian monopoles (b), for the merging case at times \(t=0,\, 0.38,\, 0.76,\, 1.13,\, 1.51\). (a) The different patches correspond to successive time steps, showing how they approach each other from their initial positions \((8,10.5)\), \((12,9.5)\). (b) Successive monopole positions.}
\label{f-patch}
\end{figure}

It is interesting to compare more quantitatively the motion of the point vortices with the actual evolution governed by the full equations. One obvious difference between the two approaches is that point vortices lack extension, in contrast to the Gaussian vortices of the full simulation. However, one may introduce a patch of point vortices over a region whose size is of the order of the Gaussian monopole, and follow their Hamiltonian dynamics. Such a comparison is presented in Fig.~\ref{f-patch}, where we chose the same initial positions and wave phase, and adjusted the point vortices strength to obtain similar time scales as in the extended vortex system. We observe that the approach of the two coherent structures follows the same pattern, including the enhancement of the asymmetry between the bottom and top vortices; both patches are strongly stretched during they approach, at \(t\approx 1.5\). Indeed, in the Charney-Hasegawa-Mima simulation we also observe that the top monopole is slightly reinforced as 
it approaches the bottom vortex, which moves faster than the top one. However a difference arises in the \(y\)-direction displacement; while in the vortex model its mean value is zero, for extended vortices there is a supplementary drift coming from the background vorticity gradient \cite{chavanis1998,chavanis2001,Schecter-2001fk}. Although this effect is responsible of a quantitative difference between the point and extended vortex cases, the determination of the appropriated range of parameters for vortex merging as determined by the point vortex model, appears to be rather robust against this effect. 

\begin{figure*} 
\centering
\includegraphics[width=0.30\textwidth]{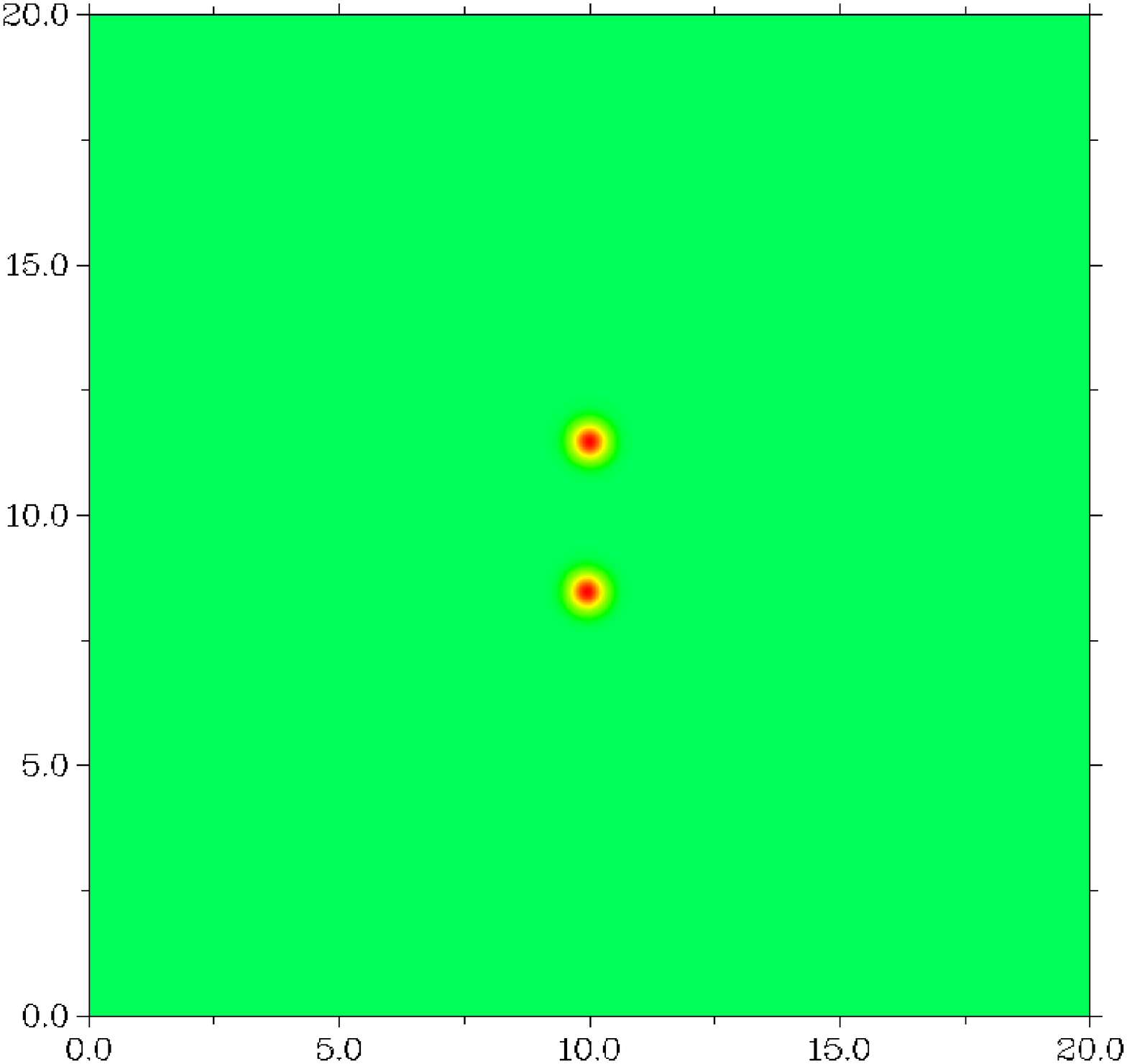}(a)\hfill%
\includegraphics[width=0.30\textwidth]{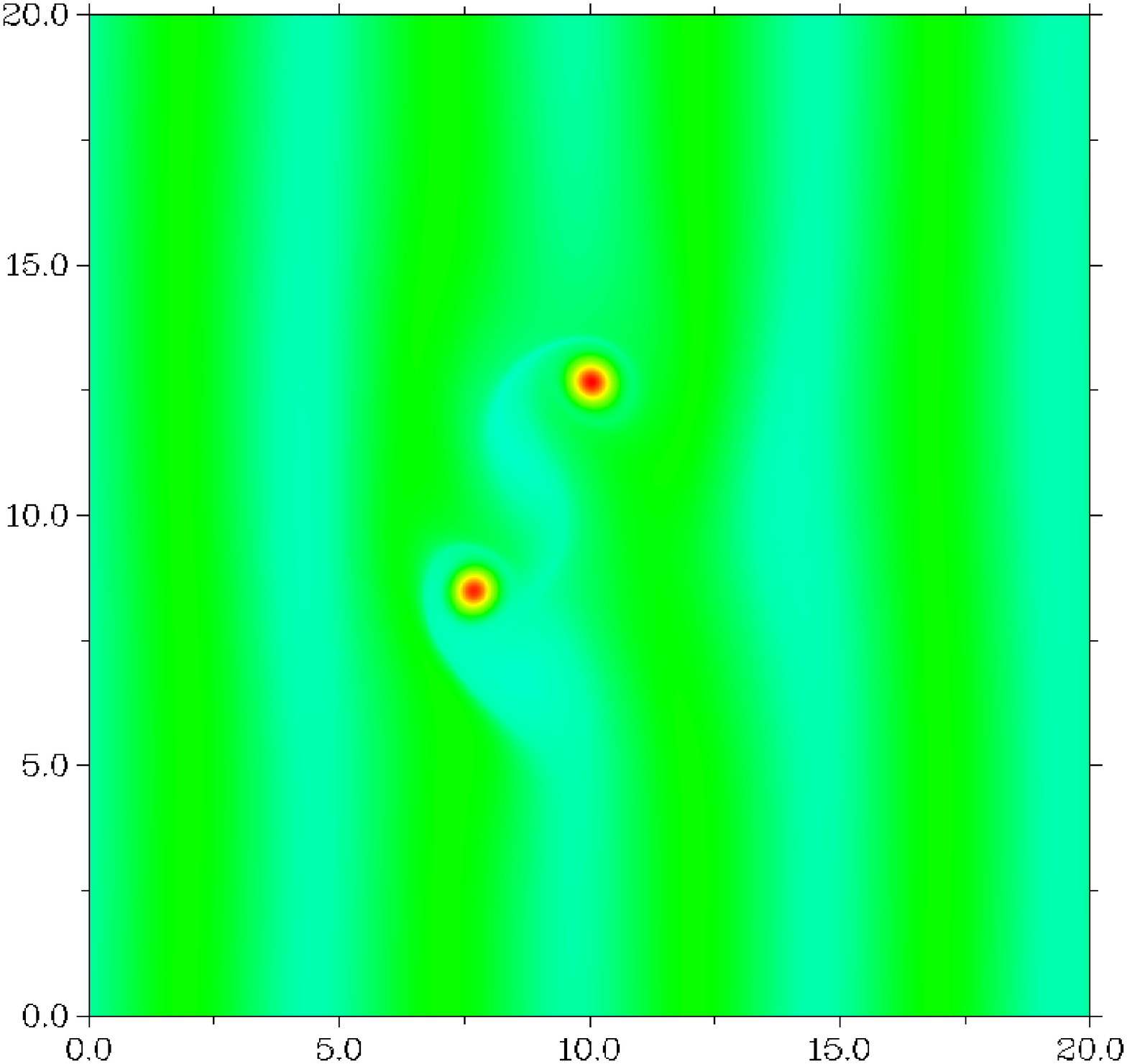}(b)\hfill%
\includegraphics[width=0.30\textwidth]{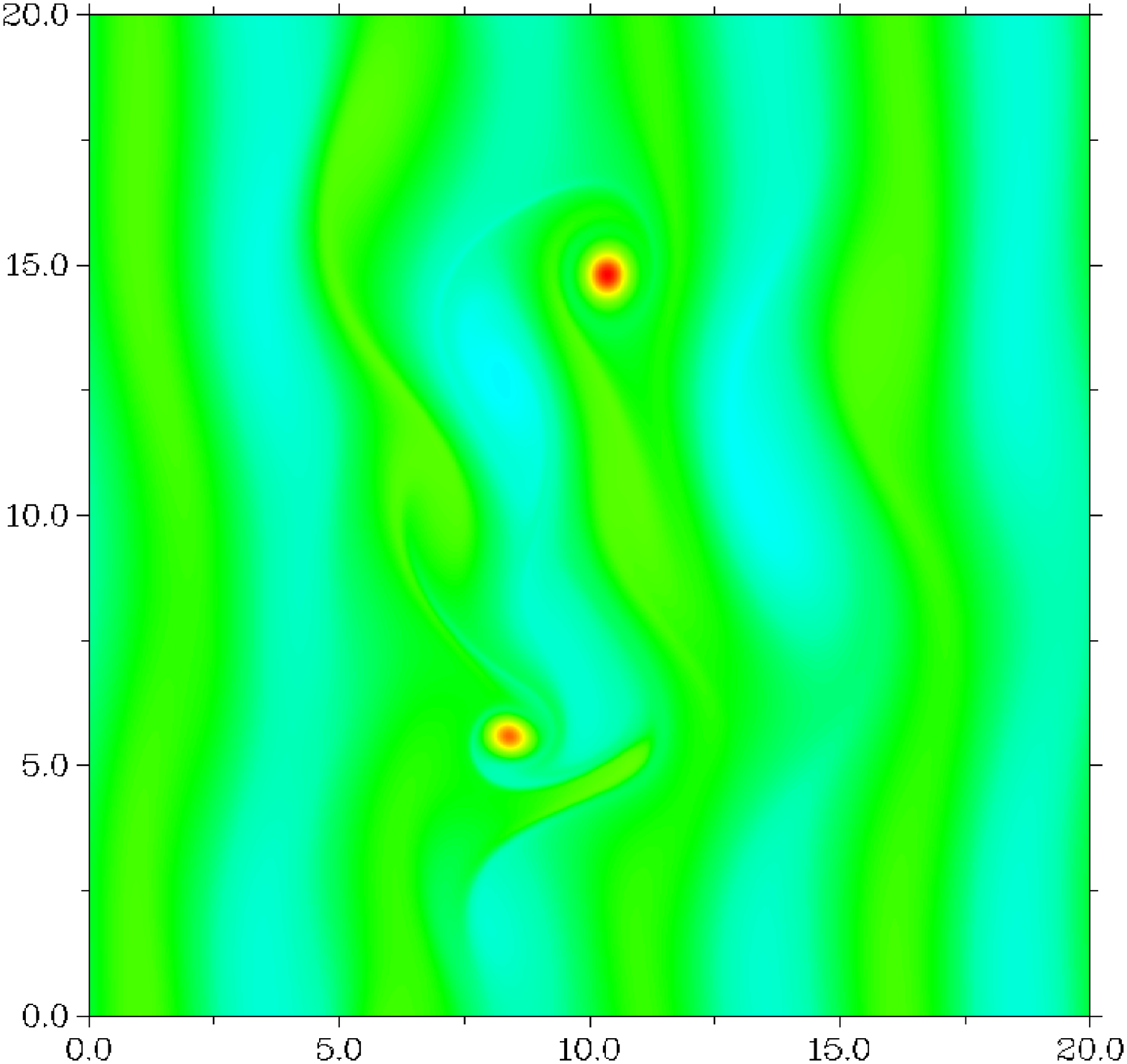}(c)\\
\includegraphics[width=0.30\textwidth]{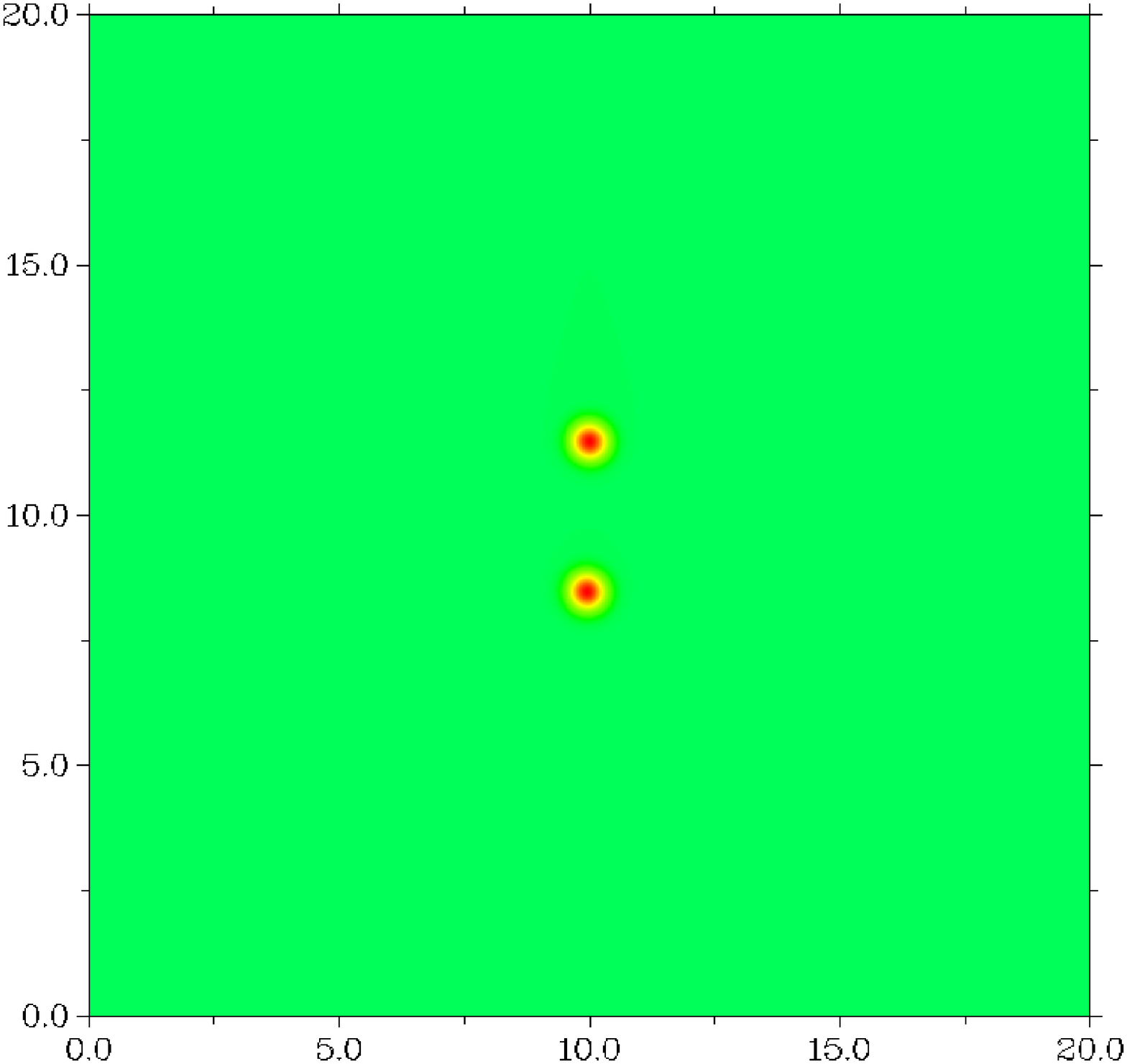}(d)\hfill%
\includegraphics[width=0.30\textwidth]{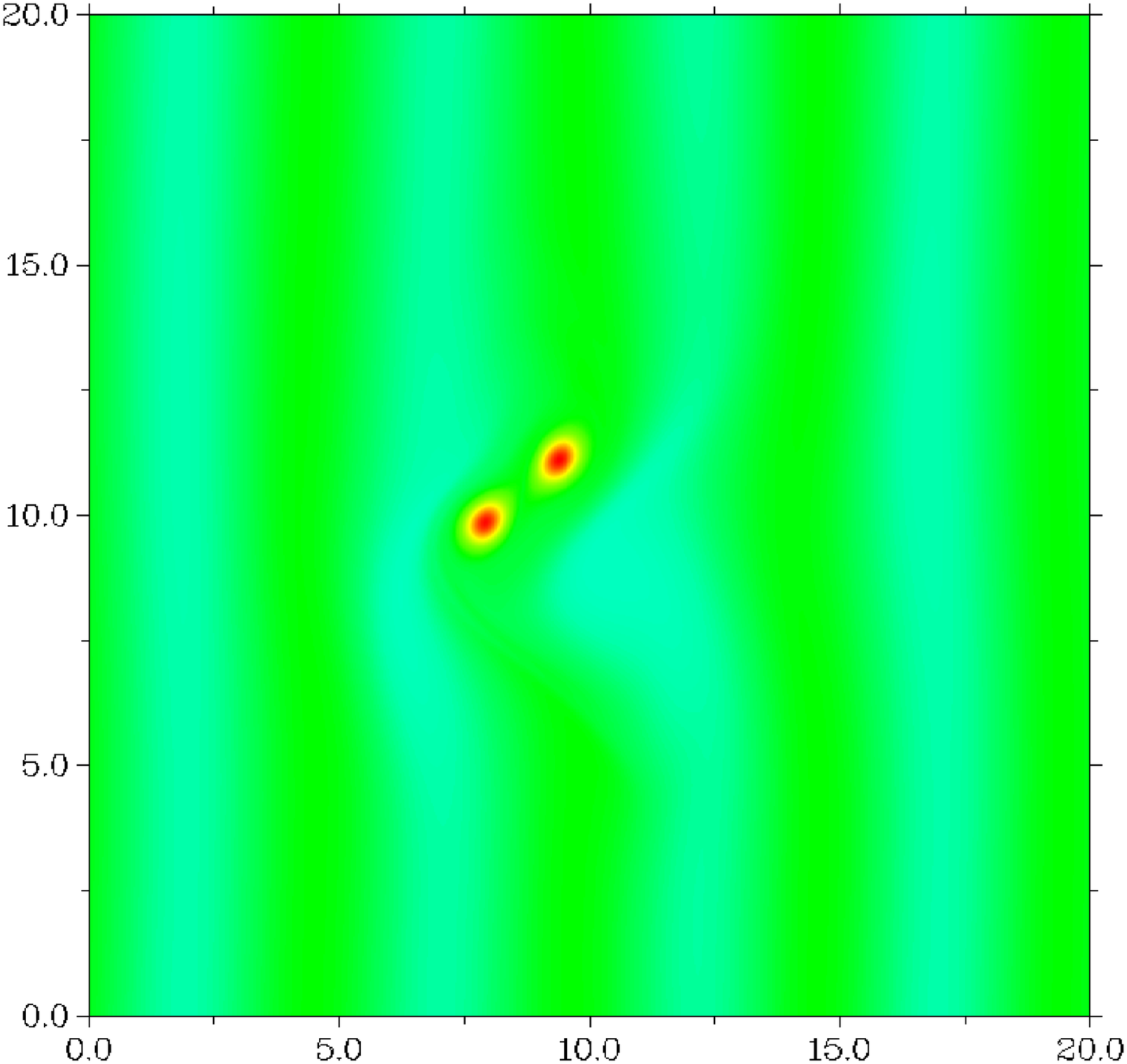}(e)\hfill%
\includegraphics[width=0.30\textwidth]{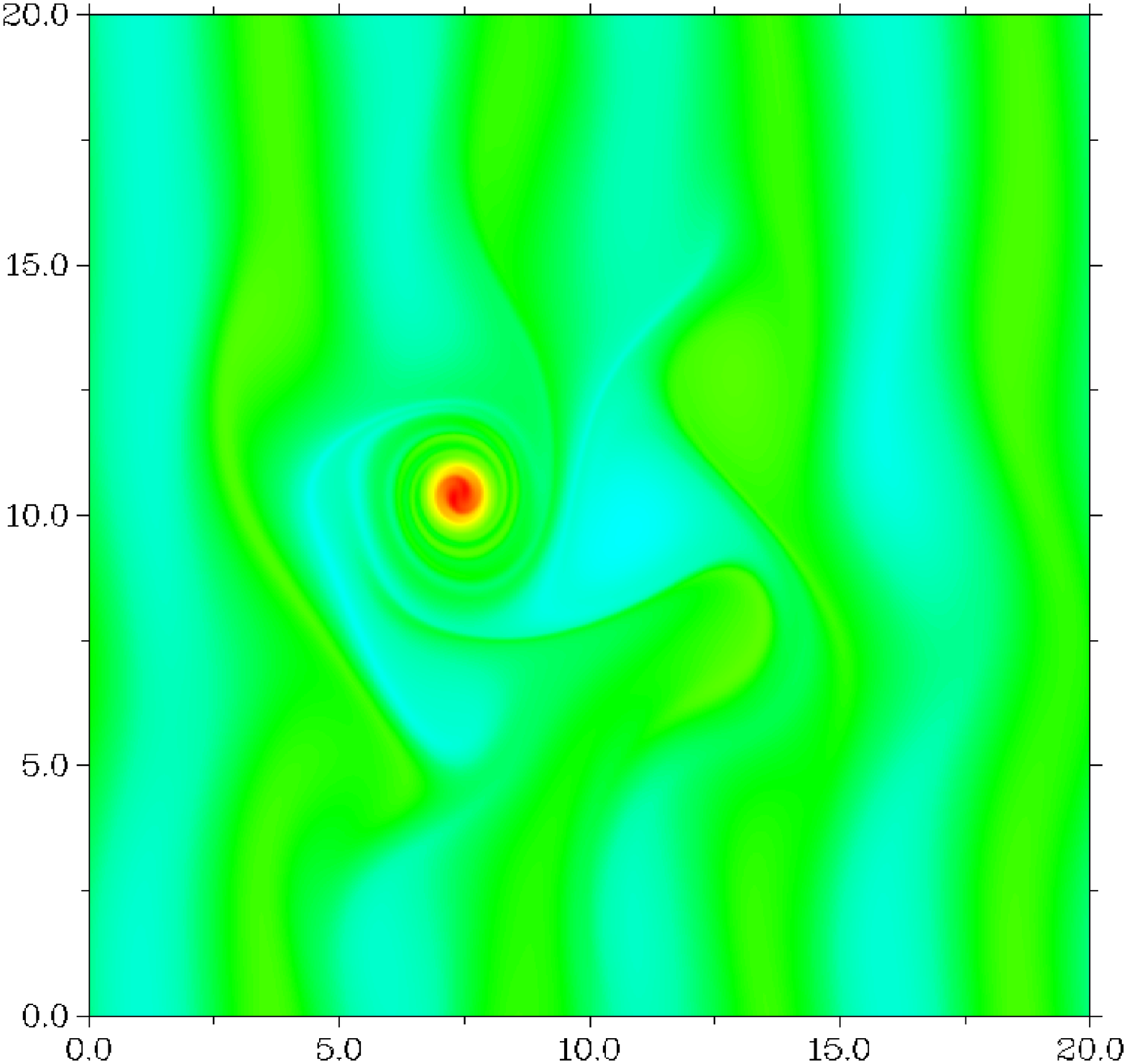}(f)
\caption{(Color online) Evolution of two gaussian monopoles with a source of longitudinal waves. Splitting (a-d) or merging (d-f) depend on the phase of the source term. (a,d), initial state; (b,e), intermediate state (\(t = 1.5\)); (c,f), final state (\( t = 3\)). Vortex and wave parameters are as in the transverse case (Fig.~\protect\ref{f-t}).}
\label{f-l}
\end{figure*}

\subsection{Longitudinal wave}

We turn out now to the study of the longitudinal wave driven system. The direction of the longitudinal wave can be in principle arbitrary. However, it is convenient to well separate the longitudinal and transverse situations in order to identify the physical mechanisms. Therefore, we take a source that generates longitudinal waves perpendicular to the propagating intrinsic transverse waves. This is the best suited situation to apply the results of the point vortex model. Although the source term added to the vorticity equation is restricted to a wave of the form \(\hat{\bm x} \sin(kx-\omega t)\), the evolution of the two equal circulation vortices can display a large variety of behaviors. We focus here, on the strong interaction of vortices induced by the wave, and leading to their separation or fusion as predicted by the point vortex model. Therefore, we take parameters similar to the ones of Figs.~\ref{f-fusi} and \ref{f-split}, for the vortices and wave intensities, positions and phase. Remarkably, we 
obtained the same kind of trajectories in the extended case, shown in Fig.~\ref{f-l}, where the vortex pair split or merge according to the phase of the longitudinal wave.

We remark 
  that the relative position of the monopoles with respect to the wave influences the behavior of their vorticity: the wave possesses its own vorticity which superposes to the monopoles one, reinforcing the monopole in the positive vorticity background and weakening the one in the negative vorticity background and causing additional drift \cite{Kiwamoto-2000vn,chavanis1998,chavanis2001,Schecter-2001fk}. This effect, in the observed regime of parameters, slightly affects the motion of the vortices during the time of their evolution, without changing qualitatively the fusion or splitting process. In addition, the vortex interaction mechanisms resulted to be much faster than the evolution of the wave vorticity, showing that neglecting as a first approximation  the influence of the vortex motion on the wave vorticity appears to be justified.

Although we chose to present results on the monopole dynamics in the general case, we note that the point vortex model predicts, in the presence of a longitudinal or transverse wave, that a dipole will follow an almost straight trajectory with a superposed oscillation, along certain directions with respect to the wave propagation. This behavior was also observed in dipole systems obeying the Hasegawa-Mima equation \cite{Makino81}.

\section{Conclusions}
\label{Sec:Con}
In this paper we introduced a simple model based on the dynamics of point vortices, in order to get an 
idea on how extended vortices and waves interact in more realistic flows. As a matter of fact, it 
turned out that this simple model was much richer in the variety of 
behaviors it could reflect, than what would be expected. Namely, it 
demonstrated how stable a dipole was, and in opposition how like-circulation vortices 
could be destroyed or merged by a wave. Comparison of the point vortex plus wave analytical results and the Charney-Hasegawa-Mima simulations, shown that the distinction between localized and extended vorticity is justified by the existence of well separated length and time scales, and is also relevant to determine, for instance, the conditions for vortex merging or splitting.

We considered in particular, the interaction of equal circulation vortices in the presence of an intrinsic drift (transverse) and of an external (longitudinal) wave. We found that, depending on the relative phase of the wave with respect to the positions of the monopoles, merging, splitting or even chaotic trajectories are possible. Using the same set of parameters, we confirmed that the point vortex model reasonably predicts the behavior of the extended vortices obeying the Charney-Hasegawa-Mima equation.

In summary, the results obtained in the case of the point vortex model and confirmed by 
numerical simulations, could 
partially explain how coherent structures can appear by the merging 
of small structures and then organize themselves into dipoles
 and monopoles, especially in systems where waves can 
be present or generated by the dynamics of the structures. We may hope that this simple model, that permits to identify the range of parameters relevant for the vortex-wave interactions, could be a help 
in  understanding the dynamics of large 
coherent structures in decaying turbulent flows.

\begin{acknowledgement}
The authors gratefully acknowledge helpful discussions with O. 
Agullo. We would also like to thank the laboratories PIIM and IRPH\'E, for hospitality and support during the early development of this work. This work was carried out within the framework the European Fusion Development Agreement and the French Research Federation for Fusion Studies. It is supported by the European Communities under the contract of Association between Euratom and CEA.
 The views and opinions expressed herein do not necessarily reflect those of the European Commission.
\end{acknowledgement}

\section{Appendix: Charney equation with source term}
\label{Sec:app}

In this appendix we briefly discuss the derivation of the Charney-Hasegawa-Mima equation in the geostrophic framework \cite{Pedlosky_book}, in order to establish its relation with (\ref{vadvection}) and the origin of the source term.

The velocity field \(\bm v\) satisfies the motion equation,
\begin{equation}
\frac{d }{d t}\bm v=
   -\frac{\nabla P}{\rho} - g\hat{\bm z} + f\bm v\times \hat{\bm z}\,,
\label{motion}
\end{equation}
where 
\[
\frac{d }{d t}=\frac{\partial}{\partial t}+\bm v\cdot\nabla\,,
\]
is the total derivative, \(\rho\) is the fluid density, \(P\) the pressure, \(g\) the acceleration of gravity, and \(f=f(y)\) the Coriolis frequency (which is a function of the latitude). The usual choice of coordinates is \(x\) in the equator east direction and \(y\) in the north direction. Note that the plasma convention, as in (\ref{mimeq}), exchange the role of the coordinates: \((x,y)\rightarrow(y,-x)\). The continuity equation writes,
\begin{equation}
\nabla\cdot\bm v=-\frac{d }{d t}\ln\frac{\rho}{\rho_0}\,,
\label{continuity}
\end{equation}
where \(\rho_0=\mathrm{const}\). The equation of state, for instance the adiabatic relation between pressure and density \(P\sim\rho^\gamma\), complete the basic equations. Applying the rotational to (\ref{motion}), one obtains the vorticity evolution equation,
\begin{align}
\frac{d }{d t}(\bm\omega + f\hat{\bm z})&+
	(\bm\omega + f\hat{\bm z})\nabla\cdot \bm v-\nonumber\\
	&(\bm\omega + f\hat{\bm z})\cdot\nabla \bm v=
		- \nabla\frac{1}{\rho}\times\nabla P \,,
\label{vorticity}
\end{align}
where \(\bm \omega=\nabla\times\bm v\). 

We consider a fluid layer at height \(H_0\) of width \(h=h(\bm x,t)\), with \(\bm x=(x,y)\). To describe the dynamics of this layer it is convenient to separate in the equation of motion (\ref{motion}), the horizontal (\(\perp\)) and vertical (\(z\)) components:
\begin{equation}
\frac{d }{d t}\bm v_\perp=
   -\frac{1}{\rho}\nabla_\perp P+ f\bm v\times \hat{\bm z}\,,
\label{motionperp}
\end{equation}
and
\begin{equation}
\frac{d v_z}{d t}=
	-\frac{1}{\rho}\frac{\partial P}{\partial z} -g\,,
\label{vertical}
\end{equation}
respectively. We also consider that in addition to the hydrostatic pressure \(\rho_0 g h\), there can be a forcing pressure \(p=p(\bm x, z, t)\) such that the total pressure writes,
\begin{equation}
P=p_0+\rho_0 g h+p\,.
\label{pressure}
\end{equation}
The extra pressure term induces a vertical flow,
\begin{equation}
\frac{\partial v_z}{\partial t}=
	-\frac{1}{\rho_0}\frac{\partial p}{\partial z}\,,
\label{verticalp}
\end{equation}
where we neglected the nonlinear convective term and replaced everywhere the density by its mean value \(\rho_0\). Moreover, from the horizontal motion equation (\ref{motionperp}), the balance of the hydrostatic pressure gradient and the Coriolis force leads to a drift velocity,
\begin{equation}
\bm v_\perp = \frac{g}{f_0}\hat{\bm z}\times\nabla h\,,
\label{driftv}
\end{equation}
where, to this order, the Coriolis frequency is replaced by its averaged value \(f_0\), and the inertia and unsteady pressure \(p\) terms, are neglected \cite{Charney-1948uq}. The total velocity field can thus be written as, 
\begin{equation}
\bm v = \bm v_\perp+\nabla \phi\,,
\label{vtotal}
\end{equation}
where the second term, that takes into account the fluctuations around the equilibrium drift velocity (\ref{driftv}), is of the same order as the fluctuating pressure term in (\ref{pressure}). Equation (\ref{vtotal}) can be compared with (\ref{vadvection}), to note that the point vortex model reduces to the advection of the localized part of the height \(h\) field, in the field of the drift velocity \(\bm v_\perp\), possibly driven in addition, by a external velocity field \(\nabla \phi\). Using this expression in (\ref{verticalp}), one obtains,
\begin{equation}
\frac{\partial}{\partial t}\phi = -\frac{p}{\rho_0}\,,
\label{phiz}
\end{equation}
which relates the pressure fluctuation with a vertical (potential) motion.

Introducing (\ref{driftv}) into (\ref{vorticity}), and neglecting the baroclinic term (which vanishes for adiabatic motions), the vorticity equation becomes,
\begin{equation}
\frac{d }{d t}(\omega+f)+
	(\omega + f)\nabla\cdot \bm v_\perp=0,,
\label{vorticityperp}
\end{equation}
where \(\omega \hat{\bm z}=\nabla \times \bm v_\perp\). Using now (\ref{continuity}), with 
\[
\ln\frac{\rho}{\rho_0} \approx 
	\frac{1}{\gamma} \ln\left(1+\frac{h}{H_0}+\frac{p}{p_0}\right)\,,
\]
neglecting \(\omega\) in the second term of (\ref{vorticityperp}), and averaging over the vertical layer, one obtains,
\begin{equation}
\frac{d }{d t}\left(\omega+f -\frac{f_0}{\gamma H_0} h\right)=f_0
	\left\langle  
		\frac{\partial}{\partial t}\frac{p}{\gamma p_0}
		+\frac{\partial v_z}{\partial z}
	\right\rangle\,,
\label{vorticityperp1}
\end{equation}
where all the quantities depend on the horizontal coordinates. The right-hand side term can be explicitly calculated using (\ref{phiz}),
\begin{equation}
\left\langle\frac{\partial}{\partial t}\frac{p}{\gamma p_0}
		+\frac{\partial v_z}{\partial z} \right\rangle= \frac{-1}{c_s^2}
		\left\langle
			\frac{\partial^2 \phi}{\partial t^2}-
			c_s^2\frac{\partial^2 \phi}{\partial z^2}
		\right\rangle=
	 \frac{S(\bm x, t)}{f_0}\,,
\label{acoustic}
\end{equation}
where \(c_s=(\gamma p_0/\rho_0)^{1/2}\) is the sound velocity, and where we defined \(S=S(\bm x, t)\), the effective source of vorticity that contains the contribution of vertical unsteady motions. The appearance of the sound speed is a consequence of the assumption on the equation of state. An equivalent equation, with a different source term, can be obtained if instead of an external force, one introduces in the energy equation a heat source, and relates the pressure fluctuations with the temperature fluctuations \cite{Lorenz-1960fk,Pedlosky_book}. It is worth noting that if \(\phi\) satisfy the wave equation, the source term reduces to the form \(S\sim k^2\phi\), where \(\bm k=(k_x,k_y)\) only contains the perpendicular component of the wavevector. Although the specific form of the source term will depend on the physical mechanism from which it originates, being proportional to \(\partial v_z/\partial z\), it naturally generate longitudinal perturbations.

Replacing \(\omega=(g/f_0)\nabla^2h\) and (\ref{acoustic}) into (\ref{vorticityperp1}), and putting \(f=f_0 + \beta y\), we finally get
\begin{equation}
\frac{d }{d t}\left(\frac{g}{f_0}\nabla^2h+\beta y-\frac{f_0}{\gamma H_0} h\right)=S(\bm x, t)\,.
\label{h-eq}
\end{equation}
In the usual quasi-geostrophic approximation, the \(S\) term is neglected, and one recovers the Charney-Hasegawa-Mima equation (\ref{CHM}), where we used the following identifications between plasma and fluid parameters: \(\psi=gh/f_0\), \(v_d=\beta \gamma H_0/f_0\), and \(\rho_s^2=\gamma H_0/f_0\), and changed axes according to the plasma convention, \((x,y)\rightarrow(y,-x)\). 

%
%
\bibliographystyle{epj}

\begin{thebibliography}{41}
\bibitem{remark}Note that in the case of unequal circulations there is an extra length scale, as both vortices rotate on different circles around the center of vorticity but the qualitative behavior of the system is  essentially not modified.
\end{thebibliography}

\end{document}